\documentclass[12pt]{article}
\usepackage{latexsym}
\usepackage{psfig}

\newcommand\cc{\hbox{\rm \bf c}}
\newcommand\rr{\hbox{\rm \bf r}}
\newcommand\Or{\hbox{\cal O}}

\begin{document}

\begin{center}
{\large \bf Chaos properties and localization in
Lorentz lattice gases}
\vskip 1cm
{C. Appert\footnote{permanent address: CNRS, LPS,
Ecole Normale Sup\'erieure, 24 rue Lhomond,
75231 PARIS Cedex 05, France}
 and M.H. Ernst\\
Instituut voor Theoretische Fysica, Univ. Utrecht\\
Postbus 80 006, 3508 TA UTRECHT, The Netherlands\\
appert@physique.ens.fr\\
(\today)
}
\end{center}
 
\vskip 0.5 cm
 
\begin{abstract}
The thermodynamic formalism of Ruelle, Sinai, and Bowen,
in which chaotic properties of dynamical systems are
expressed in terms of a free energy-type function $\psi(\beta)$,
is applied to a {L}orentz {L}attice {G}as,
as typical for diffusive systems with static disorder.
In the limit of large system sizes, the mechanism and
effects of localization on large clusters of scatterers
in the calculation of $\psi(\beta)$ are elucidated and supported 
by strong numerical evidence.
Moreover it clarifies and illustrates a previous
theoretical analysis \protect{\cite{appert96c}}
of this localization phenomenon.
\end{abstract}
 

\vskip 2.5 cm
\section{Introduction}
 
Many nonlinear physical problems involve
a complicated discrete distribution function $\{p_i\}$.
As it may vary in a very irregular way with $i$,
it would be attractive to replace the $p_i$'s by a smooth function
containing the same information about the structure of
the distribution.
One way to do it is to associate with the distribution
a whole set of so--called ``escort distributions''
\cite{beck_s93} defined as
$P_i = (p_i)^\beta/Z(\beta)$, where $Z(\beta)$ is
a partition function $Z(\beta) \equiv \sum_i (p_i)^\beta$.
The parameter $\beta$ allows one to scan the structure of
the initial distribution. Large $\beta$ values enhance
the most probable trajectories whereas negative
$\beta$'s focus on the least probable trajectories
(we impose that $(p_i)^\beta = 0$ if $p_i = 0$ so that
our definitions still hold for negative $\beta$'s).
By analogy with thermodynamics where $\beta$ would be
an inverse temperature, a free energy---like function
$\psi(\beta)$
is introduced, which is related to the logarithm of
the dynamic partition function $Z(\beta)$.
 
This formalism introduced by Ruelle, Sinai and Bowen
 \cite{ruelle78} is called {\em thermodynamic formalism}.
It has been successfully applied for instance to
multifractals \cite{feigenbaum_p_t89}.
 
The present paper deals with another frame of application
of this formalism, i.e. the chaotic properties
of dynamical systems.
For a given map (we will assume hyperbolicity
in order to ensure good ergodicity properties),
the phase space is partitioned into cells.
Each sequence of cells explored by a trajectory
$(r_o, \cdots, r_t)$ in $t$ time steps is one
point $\Omega_t$ of the {\em dynamical phase space}.
We shall refer to it as a trajectory over
$t$ time steps\footnote{Mathematically,this could also be
formulated in terms of cylinder sets, i.e. the set
of initial conditions that follow the trajectory $\Omega_t$
during the first $t$ time steps.}.
 
The thermodynamic formalism will be applied
to the distribution $P(\Omega_t)$, where $\Omega_t$
replaces the previous subscript $i$.
The partition function reads now
$Z(\beta,t) = \sum_{\Omega_t} P^\beta(\Omega_t)$.
Note that each $\Omega_t$ is counted with equal weight,
i.e. the calculation does not require the a priori
knowledge of the invariant measure.
The free energy is now\footnote{For systems with escape,
$Z(\beta,t)$ vanishes exponentially.}
defined as
$\psi(\beta) = \lim_{t \rightarrow \infty}
\frac{1}{t} \ln Z(\beta)$. In this context,
it is called {\em Ruelle pressure} or
{\em topological pressure}.
It contains information about the
dynamics of the system, which can be extracted
again by varying the control parameter $\beta$.
For example, for an open system, the escape rate $\gamma$
is such that $\psi(1) = -\gamma$, and
$\psi(\beta)$ vanishes when $\beta$ is
equal to the fractal dimension of the repeller
(set of trajectories which never escape).
 
It has already been found in a more general frame
that some non analyticity
of the Ruelle pressure may occur at certain $\beta$
values \cite{beck_s93,gaspard_b95,feigenbaum_p_t89}.
They are called phase transitions in analogy
with thermodynamics.
 
In this paper, we study the influence of disorder
on the structure of the Ruelle pressure.
We have applied the
thermodynamic formalism to systems with
static disorder and have found very peculiar features
\cite{appert96b}.
In the limit of infinite systems,
and for almost all $\beta$ values, the Ruelle pressure
becomes completely determined by trajectories
localized on rare fluctuations of the disorder.
The global structure of the disorder ---in particular
the density of impurities--- becomes irrelevant.
Depending on $\beta$, the relevant trajectories are
not always localized onto the same type of fluctuations.
The crossovers between the different localization regimes
are characterized by non-analytical points in the
Ruelle pressure.
 
For finite but large systems, the Ruelle pressure is again
analytic. However for most $\beta$'s it is still
dominated by localized trajectories. Delocalized
trajectories become relevant only in some limited
$\beta$ regions which shrink to points as the system
size tends to infinity.
 
To be more precise, we have developed these ideas
by studying a particular model, the {\em Lorentz Lattice Gas}
(LLG).
A Lorentz gas consists of a moving light particle scattered
among fixed heavy particles.
It may be considered as a simple model for diffusion,
electric conductivity or flow inside a porous medium.
In the past decades, it has been successfully used to study
the connection
between the irreversible behavior of fluids and their
chaotic properties at a microscopic level.
For example, P. Gaspard and G. Nicolis \cite{gaspard_n90}
have shown that for an open system,
the diffusion coefficient can be expressed
in terms of the Kolmogorov-Sinai entropy and the
sum of the positive Lyapunov exponents for
non escaping trajectories.
The name Lorentz lattice gas is used when
the light particle is, for reasons of simplicity,
constrained to move on a (cubic) lattice, with
scatterers located at the nodes --or sites.
 
The localization processes occurring when
the thermodynamic formalism is applied to such a system,
have already been discussed in an extensive theoretical
analysis \cite{appert96c}.
The present paper contains the numerical counterpart
and demonstrates the validity of
our analytical results.
It clarifies the mechanism of localization in large systems
by explicitly calculating the topological pressure
for specific non-random configurations in section
\ref{sect_special}.
In section \ref{sect_dh} an exact expression for the
Hausdorff dimension of the repeller is obtained
for an open LLG.
Moreover, numerical studies allow us to find a good estimate for the
Ruelle pressure for large but finite systems.
As an intermediate result, the distribution
for the largest cluster of scatterers
in random configurations (section
\ref{sect_largestsize}) is estimated theoretically,
and measured through numerical simulations.
 
\section{Thermodynamic formalism}
\label{sect_models}

First we give a more precise definition of our model.
A number of $N$ fixed scatterers are randomly
placed with a probability
$\rho$ on the
$L$ sites of a finite one-dimensional lattice ${\mathcal D}$,
having either periodic or absorbing boundaries.
The presence of a scatterer at site $\rr = \{1,2,\cdots,L\}$
is indicated by the boolean variable $\hat{\rho}(\rr)$,
equal to 1 (0) if the site $\rr$ is occupied (empty),
with $\langle \hat{\rho} \rangle = \rho = N/L$.

A light particle, moving on the lattice,
is specified at time $t$ ($t=0, 1, 2, \cdots$)
by a state $x_t = \{\rr_t, \cc_{it}\}$,
where its position $\rr$ is a site on the lattice and
its velocity, $\cc_i = \pm 1$, connects site $\rr_t$ to one
of its nearest neighbors. 
Let $\phi(x,t) = \phi_i(\rr,t)$ be the probability to find
the moving particle in state $x=\{\rr,\cc_i\}$.

The time evolution of the particle from time $t$
to $t+1$ consists
of a possible collision followed by propagation.
The collision is defined by the following rules :
if there is no scatterer at $\rr_t$, then the particle
moves ballistically, i.e. $\rr_{t+1} = \rr_t + \cc_i$.
This is expressed in terms of $\phi_i$ as :
\begin{equation}
\phi_i(\rr+\cc_i,t+1) = \phi_i(\rr,t).
\label{noscat}
\end{equation}
If there is a scatterer at $\rr_t$, i.e. if
$\hat{\rho}(\rr_t) = 1$, then the velocity
of the particle is reversed with probability $q$
($\cc^\prime_{it} = -\cc_{it}$)
or left unchanged with probability $p$
($\cc^\prime_{it} = \cc_{it}$), with $p + q = 1$.
In the propagation step the particle moves over one
lattice distance in the direction of its post-collision
velocity, i.e. $\rr_{t+1} = \rr_t + \cc^\prime_{it}$.
Then, it hops from one site to the next one
in the direction of its velocity $\cc_i$.
Again, the corresponding evolution of $\phi_i$ is
given by :
\begin{equation}
\phi_i(\rr+\cc_i,t+1) = a \phi_i(\rr,t) + b \phi_{-i}(\rr,t)
\label{withscat}
\end{equation}
with $a = p$ and $b = q = 1-p$ (the reason for this
notation will become clear later).
More generally, we define site dependent transition
probabilities
\begin{eqnarray}
\hat{a}(\rr) & = & a \hat{\rho}(\rr) + 1 - \hat{\rho}(\rr)
\nonumber \\
\hat{b}(\rr) & = & b \hat{\rho}(\rr).
\label{}
\end{eqnarray}
They depend on the precise configuration of scatterers
under consideration.

The probability $\phi(x,t)$ evolves with
time according to a Chapman--Kolmogorov (CK) equation
with site dependent transition probabilities,
obtained by combining equations (\ref{noscat}) and
(\ref{withscat}) :
\begin{equation}
\phi_i(\rr + \cc_i,t+1) = \hat{a}(\rr) \phi_i(\rr,t)
+\hat{b}(\rr) \phi_{-i}(\rr,t),
\label{CK1}
\end{equation}
and more formally
\begin{equation}
\phi(x,t+1) = \sum_y w(x|y) \phi(y,t).
\label{CK2}
\end{equation}
The transition matrix $w(x|y)$ represents the probability
to go from state $y=\{\rr^\prime, \cc_j \}$ to state
$x=\{\rr,\cc_i\}$, and is given by
\begin{equation}
w(x|y) =
\delta(\rr - \cc_i, \rr^\prime )
\left[ \delta_{ij} \hat{a}(\rr^\prime) +
\delta_{i,j+1} \hat{b}(\rr^\prime) \right].
\label{defw}
\end{equation}
In the case of absorbing boundaries (open system),
boundary states
 $y=\{ \rr^\prime, \cc_j \} = \{1,+\}$ and
$\{L,-\}$ referring to a particle entering the domain
${\mathcal D}$ are excluded from the $y$--summation.
This is equivalent to imposing the absorbing boundary
conditions (ABC)
\begin{equation}
\phi_+(1,t) = \phi_-(L,t) = 0.
\end{equation}
In the case of periodic boundaries (closed system),
we impose the periodic boundary conditions (PBC)
\begin{equation}
\phi_i(\rr+L,t) = \phi_i(\rr,t).
\end{equation}
 
The transition matrix satisfies the normalization relations
\begin{equation} \label{a7}
\sum_x w(x|y) \left\{\begin{array}{cc}
= 1 & \mbox{(closed)} \\
\leq 1 & \mbox{(open)}.
\end{array}
\right.
\end{equation}
The inequality sign in Eq.(\ref{a7}) for {\em open} systems refers to the
case where $y = \{ \rr,\cc_i \} $ denotes a
state at a boundary site $\rr$ with non--entering
velocity (boundary states with entering velocity do not occur).
Indeed, the sum over $x$
excludes states where the particle has escaped from
the domain ${\mathcal D}$. Hence the probability for remaining inside the
domain decreases when the particle finds itself on a boundary site.

The special case $\rho=1$ is called
a Persistent Random Walk (PRW) where $\hat{a}(\rr) = p$
and $\hat{b}(\rr) = q$ for all $\rr$.
Then the moving particle
simply undergoes a random walk with correlated jumps.

For later analysis it is convenient to also have a different
representation of the {CK}--equation. It can be obtained
by eliminating the probabilities $\phi_i(\rr,t)$ at
non--scattering sites with the help of Eq. (\ref{noscat}).
Let $r_k$ ($k=1,2,\cdots,N$) be the position of the
$k$-th scatterer and $R_k = r_{k+1} - r_k$ the free
interval between scatterers.
For ABC we define in addition $R_o = r_1 - 1$
and $R_N = L - r_N$. Then it is straightforward
to show that the scattering amplitudes
$U_i(k,t) = \phi_i(r_k,t)$ (probability at scattering
site $r_k$) satisfy the closed set of equations
\begin{eqnarray}
U_+(k+1,t+R_k) & = & a U_+(k,t) + b U_-(k,t) \nonumber\\
U_-(k-1,t+R_{k-1}) & = & b U_+(k,t) + a U_-(k,t)
\label{scatampl}
\end{eqnarray}
with the boundary conditions
\begin{eqnarray}
&U_i(N+k,t) =  U_i(k,t) & \mbox{(PBC)} \nonumber\\
&U_+(1,t) = U_-(N,t) = 0 & \mbox{(ABC)}.
\label{scatbc}
\end{eqnarray}
Note that Eq.(\ref{scatampl}) only depends on the
set of {\em random} intervals $\{R_k\; |\; k=0,1,2,\cdots,N\}$
(the coefficients $a = p$ and $b = q$ are {\em sure}
variables),
whereas the {CK}-equation (\ref{CK1}) depends on the set of random
transition rates
$\{\hat{a}(\rr),\hat{b}(\rr)\; |\; \rr=1,2,\cdots,L\}$.
It will appear below that the basic chaos properties can
be expressed in terms of the largest eigenvalue $\Lambda$
and the corresponding left and right eigenvectors
$v(x)=v_i(\rr)$ and $u(x)=u_i(\rr)$ of the non-symmetric matrix $w$
appearing in the CK--equation (\ref{CK2}) :
\begin{eqnarray}
w u & = & \Lambda u \nonumber \\
v w & = & \Lambda v.
\label{lr_eigv}
\end{eqnarray}
Then, any solution $\phi(x,t)$ of the CK--equation
approaches $\Lambda^t u(x)$ for $t \rightarrow \infty$.
For our purpose it is again more convenient to deal with
Eq.(\ref{scatampl}) and determine only the components
$U_i(k)$ of eigenvectors at the scattering sites, i.e.
$U_i(k,t) = \Lambda^t U_i(k)$
and the eigenvalue equation
follows from (\ref{scatampl}), i.e.
\begin{eqnarray}
\Lambda^{R_k}U_+(k+1) & = & a U_+(k) + b U_-(k) \nonumber\\
\Lambda^{R_{k-1}}U_-(k-1) & = & b U_+(k) + a U_-(k),
\label{scat_eig_eq}
\end{eqnarray}
where the components $U_i(k)$ satisfy the boundary
conditions (\ref{scatbc}).

To describe the thermodynamic formalism we introduce
a dynamical phase space consisting of all possible sequences
$\Omega_t = \{x_1,x_2,\cdots,x_t\}$, which represent an
allowed (i.e. in the ABC case, non--escaping from domain ${\mathcal D}$)
of the moving particle visiting the state
$x_\tau = \{ \rr_\tau, \cc_\tau\}$ at the $\tau$--th time step.
The probability $P(\Omega_t | x_o)$ on $\Omega_t$,
given that the moving particle is in $x_o$ at $\tau = 0$,
is given by the multi--time distribution function
\begin{equation}
P(\Omega_t | x_o) = \Pi_{\tau = 0}^{t-1} w(x_{\tau+1}|x_\tau)
\end{equation}
on account of the CK--equation (\ref{CK2}).

The dynamic partition function is then introduced as a sum
over state $\Omega_t$ in this dynamical phase space :
\begin{eqnarray}
Z(\beta,t|x_o) & = &
\sum_{\Omega_t} [P\left( \Omega_t|x_o\right)] ^\beta
\nonumber\\
& = & \sum_{x_1 \cdots x_\tau}
\Pi_{\tau = 0}^{t-1} w(\beta; x_{\tau+1}|x_\tau)
\label{defZ}
\\
& = & \sum_y w^t(\beta; y|x_o). \nonumber
\end{eqnarray}
In analogy with the methods of equilibrium statistical
mechanics, there is an inverse temperature--like
variable $\beta$, which allows one to scan the structure
of the probability distribution for $\Omega$.
On the second line of (\ref{defZ}) we have introduced
the pseudo--transfer matrix, as
\begin{equation}
w(\beta; x|y) = \left(w(x|y)\right)^\beta
\end{equation}
and $w^t(\beta)$ denotes the $t$--th power of
matrix $w(\beta)$.
The largest eigenvalue $\Lambda(\beta)$
and corresponding left and right eigenvectors
$v(\beta,x)$ and $u(\beta,x)$ of $w(\beta; x|y)$
are defined analogously to (\ref{lr_eigv}),
where $a$ and $b$ in equations
(\ref{withscat})--(\ref{scat_eig_eq})
take the values :
\begin{equation}
a = p^\beta; \;\;\; b = q^\beta.
\end{equation}

As the system is ergodic \cite{ernst_d95}, $Z(\beta,t|x_o)$
does not depend
on the initial condition $x_o$ in the long time limit,
for almost all configurations of scatterers.
We already mentioned that
$Z(\beta,t|x_o)$ vanishes for open systems
as time tends to infinity.
More precisely, for large times, the sum (\ref{defZ})
is dominated by the largest eigenvalue $\Lambda(\beta)$
of the pseudo transfer matrix
$w(\beta; x|y)$, which is
non--degenerate for ergodic systems.
More explicitly, we use the spectral decomposition of
$w^t(\beta)$, i.e.
\begin{equation}
w^t(\beta;x|y) = v(\beta,x) \left[\Lambda(\beta)\right]^t
u(\beta,y) + \sum_{n\neq 0} v_n(\beta,x) 
\left[\Lambda_n(\beta)\right]^t u_n(\beta,y),
\end{equation}
where $\Lambda_n < \Lambda \;\;$ for all $n \neq 0$.
Thus the second term decays exponentially faster
than the first one, and we obtain for large $t$ :
\begin{equation}
Z(\beta,t|x_o) \simeq [\Lambda(\beta)]^t
\sum_y v(\beta,y)u(\beta,x_o).
\label{zlambdat}
\end{equation}
It should be noted that the partition function
depends on the configuration of scatterers under
consideration.

The Ruelle or topological pressure $\psi(\beta,\rho)$
is defined as the infinite time limit of the logarithm
of $Z$ per unit time step,
in a way similar to the definition of the free energy
per particle in the canonical ensemble
in the thermodynamic limit, i.e.
\begin{equation}
\psi\left( \beta, \rho \right) =\lim_{t\rightarrow \infty
}\frac 1t
\left\langle \ln Z\left( \beta ,t|x_o\right) \right\rangle
_{\rho},
\label{a10}
\end{equation}
where $\langle \ldots \rangle_{\rho}$
indicates not only an ensemble average
over all initial conditions, but also over all
realizations of the disorder, i.e. all configurations
of scatterers.
Again, it can be expressed in terms of the largest
eigenvalue of the matrix $w(\beta)$ as
\begin{equation} \label{a11}
\psi(\beta,\rho) = \langle\ln \Lambda(\beta) \rangle_\rho,
\end{equation}
where we have taken the infinite time limit inside the
configurational average.
 
For some specific $\beta$ values, the Ruelle pressure
has an explicit physical meaning :
the {\em positive} Lyapunov exponent
is $\lambda = - \psi^\prime (1)$;
the escape rate for open systems is
$\gamma = - \psi (1)$;
the Kolmogorov-Sinai entropy $h_{KS}$
follows from the generalization of Pesin's theorem
and yields
$h_{KS} = \psi(1) - \psi^\prime (1)$;
the topological entropy $h_T$ satisfies $h_T = \psi (0)$;
the Hausdorff dimension $d_H$ of the repeller
(the set of trajectories that never escape)
for an open system is the zero--point of the Ruelle pressure,
i.e. $\psi(d_H) = 0$.
A prime in the above formulas denotes a $\beta$-derivative.
 
For an open system, the transition matrix
$w$ is not stochastic, i.e. its
largest eigenvalue is strictly smaller than one.
Due to the loss of trajectories at each time step,
the eigenvectors for $\beta=1$ decay according to
\begin{eqnarray}
w u & = & \exp (-\gamma) u \nonumber \\
v w & = & \exp (-\gamma) v.
\label{lr_eigv_gam}
\end{eqnarray}
In order to obtain the invariant vector $\pi$,
the eigenvectors have to be rescaled at each time step,
as explained in Ref. \cite{gaspard_d95}.
A new transition matrix $\Pi$ is defined as
\begin{equation}
\Pi(x | y) = \exp(\gamma) w(x | y)\frac{v(x)}{v(y)}.
\end{equation}
This matrix is stochastic, i.e. its largest eigenvalue
is equal to $1$ associated to a left eigenvector $\chi(x)=1$,
as can be seen from :
\begin{equation}
\sum_x \chi(x) \Pi(x | y) = \sum_x \Pi(x | y) = 1.
\end{equation}
The corresponding right eigenvector is the invariant vector
\begin{equation}
\pi(x) = \frac{v(x) u(x)}{\langle u | v \rangle},
\label{defpi}
\end{equation}
where it can be verified that
\begin{equation}
\sum_y \Pi(x|y) \pi(y) = \pi(x).
\end{equation}
With definition (\ref{defpi}), $\pi$ is normalized.
The invariant vector $\pi$ gives the probability
of finding the particle on a given site and with a given
velocity, {\em provided} it has not escaped after
infinite time.

\section{Chaos properties of special configurations}
\label{sect_special}

\subsection{Mean--field configurations}

\label{sect_PRWconf}

In the subsequent subsections we develop a theoretical
understanding of some typical properties of configurations
of scatterers, which are relevant for describing the
dynamic partition function and Ruelle pressure in large
systems.

\begin{figure}
\centerline{\psfig{file=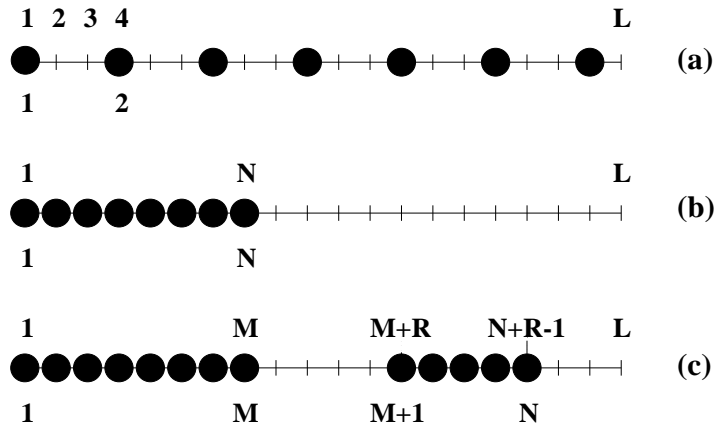,width=3.5in}}
\caption{\em Three examples of configurations with $N$
scatterers on a lattice of size $L$:
(a) mean--field configuration; (b) one solid cluster;
(c) two solid clusters of size $M$ and $\overline{M}=N-M$.
Above (below) each line of sites,
we indicate the labeling of sites (scatterers).}
\label{fig_conf}
\end{figure}

This will be done by a theoretical analysis of the largest
eigenvalue of a number of relevant configurations.
First we discuss in subsection \ref{sect_PRWconf}
mean field--type configurations, relevant for $\beta$
close to unity, as discussed in reference \cite{ernst95}.
Here the {\em fluctuations} in the lengths of the
interval between scatterers are {\em small}.
In the next subsections we study configurations with an
increasing number of {\em solid} clusters of scatterers
(regions of density $\rho=1$),
separated by regions free of scatterers (voids),
as illustrated in figure \ref{fig_conf}.
This is helpful in order to understand the mechanism of
localization.
Large voids have the tendency to divide the system
in independent subsystems with a higher density of scatterers.
We will show that only the subsystem containing the
largest cluster is relevant for determining the dynamic
partition function.

We start by considering periodic arrays of scatterers
with a constant free interval $R_k = R$ for $k=1,2,\cdots,N$
which corresponds to a PRW
on an ordered lattice with lattice distance $R$.
The eigenvalue equation (\ref{scat_eig_eq}) can be solved
by making the ansatz $U_i(k) = A_i \exp(\imath q k)$.
By setting the resulting secular determinant in
(\ref{scat_eig_eq}) equal to zero, one finds for the largest
eigenvalue $\Lambda$ :
\begin{eqnarray}
\left[\Lambda(\beta)\right]^R & = & a \cos q +
\left[b^2 - a^2 \sin^2 q \right]^{1/2} \nonumber\\
& \simeq & (a+b)
\left\{1-\frac{a}{2b} q^2 + \cdots \right\}
\;\;\;\;\mbox{(q small)}.
\label{lambdaABC}
\end{eqnarray}
The wave number $q$ has to be determined from
the boundary conditions (PBC or ABC).
In the PBC case the allowed $q$--vectors follow from
$U_i(k+N) = U_i(k)$, so that $q = 2\pi n/N$ with
$n = 0,\pm 1,\pm 2,\cdots$.
Consequently, the wavenumber $q=0$ yields the largest
eigenvalue,
$\left[\Lambda(\beta)\right]^R=a+b$.

Next consider the {\em open} system, with ABC
($\;U_+(1) = U_-(N) = 0\;$). The eigenvector
$U_i(k)$ is a special linear combination
of $\exp(\imath q k)$
and $\exp(-\imath q k)$,
i.e.
\begin{eqnarray}
U_+(k) & = & A \sin q(k-1) \nonumber\\
U_-(k) & = & B \sin q(N-k).
\label{ev_sin}
\end{eqnarray}
To determine the allowed $q$--value we substitute
equation (\ref{ev_sin}) into the equations (\ref{scat_eig_eq}),
taking $k=1$ for the first one and $k=N$ for the second one.
This yields :
\begin{eqnarray}
\Lambda^R A \sin q & = & bB \sin q(N-1) \nonumber\\
\Lambda^R B \sin q & = & bA \sin q(N-1).
\label{eq38}
\end{eqnarray}
The ratio of the two equations yields $A = \pm B$.
As the components (\ref{ev_sin}) of the largest
eigenvector of the positive matrix $w(\beta)$ need
to be positive (they represent probability densities),
it follows that $A = B = 1$.
Substituting expressions (\ref{ev_sin}) in the first
equation of (\ref{scat_eig_eq}) with $k=2$, and
using equation (\ref{eq38}) to eliminate $\Lambda^R$ from
the left hand side,
we obtain a transcendental equation for the wavenumber $q$
\begin{equation}
a \sin q = b \sin (q N)
\end{equation}
valid for $N = 1,2,\cdots$ with ABC.
For {\em large} $N$, the {\em smallest} root
of this equation is close to $q\simeq \pi/N$;
so we set $q N = \pi - \delta q$ such that $\sin (qN) =
\sin (q\delta)$ and find for small $q$ that $\delta = a/b$.
This yields for the ABC case the smallest allowed wavenumber :
\begin{equation}
q = \frac{\pi}{N+\delta} = \frac{\pi}{N+a/b}
\label{qapprox}
\end{equation}
and the largest eigenvalue :
\begin{equation}
\left[\Lambda(\beta)\right]^R = (a+b)\{1-\frac{a}{2b}
\left(\frac{\pi}{N+a/b}\right)^2 + \Or (N^{-4}) \}
\label{lambdaR}
\end{equation}
From the summary below equation (\ref{a11}), all chaos quantities
for periodic arrays of scatterers can be calculated
from (\ref{lambdaR}).
The {\em mean field theory} for the LLG, discussed in
\cite{ernst95}, follows from this result
by setting $R$ equal to the average free interval
length (mean free path) $R = L/N = 1/\rho$
and the resulting Ruelle pressure follows from
(\ref{a11}) and (\ref{lambdaR}) as
\begin{equation}
\psi^{RW}(\beta,\rho) = \rho \ln (a+b) -
\frac{a}{2b\rho}
\left(\frac{\pi}{L+a/(b \rho)}\right)^2 + \Or (L^{-4})
\label{lambdaMFII}
\end{equation}
as obtained in \cite{ernst95}.

\subsection{PBC---Configurations with a void}

\label{sect_PBCvoid}

An analysis of the eigenvalue $\Lambda(\beta)$ for
a configuration with a {\em single} void of width $R$
(i.e. $R-1$ empty sites) in the PBC and ABC case
(see figure \ref{fig_conf}b)
provides essential insights for dealing with more
complex distribution of scatterers. In this subsection,
PBC's are treated, in the next one ABC's.

We start with a perturbative calculation for large $R$.
Recall that the Hausdorff dimension $d_H$ of the repeller
is defined through $\psi(d_H) = \ln \Lambda(d_H) = 0$,
and consider first 
$\beta < d_H$, so that $\Lambda(\beta)>1$.
The position of the left--most scatterer of the cluster
is chosen as site number 1.
The equations (\ref{scat_eig_eq})
for the scattering amplitudes read in this case
\begin{eqnarray}
\Lambda^R U_+(1) & = & a U_+(N) + \epsilon b
\left[U_-(N)/\epsilon\right]
\nonumber \\
\Lambda U_-(1) & = & b U_+(2) + a U_-(2)
\nonumber \\
\Lambda U_+(2) & = & \epsilon a \left[U_+(1)/\epsilon\right]
 + b U_-(1)
\nonumber \\
\Lambda U_-(2) & = & b U_+(3) + a U_-(3)
\nonumber \\
& \cdots & \label{syst_pbcspecial} \\
\Lambda U_-(N-1) & = & b U_+(N) + \epsilon a \left[U_-(N)/\epsilon\right]
\nonumber \\
\Lambda U_+(N) & = & a U_+(N-1) + b U_-(N-1)
\nonumber \\
\Lambda^R U_-(N) & = & \epsilon b \left[U_+(1)/\epsilon\right]
 + a U_-(1)
\nonumber
\end{eqnarray}
where the $\epsilon (1/\epsilon)$ factor has been
introduced to clearly display the structure of the following
perturbative calculation.
For a {\em large} void of length $R$ the solution of
(\ref{syst_pbcspecial}) for a configuration in a {\em closed} system
of length $L = N+R-1$ is expected to look like that for
an {\em open} system of length $N$ with $N$ scatterers,
studied in subsection \ref{sect_PRWconf}.
This can be inferred from the first and last equation
in (\ref{syst_pbcspecial}), where
$\epsilon \equiv \Lambda^{1-R}$ is a small quantity
for large $R$ : the eigenvector components
$U_+(1)$ and $U_-(N)$, corresponding to ``entering''
velocities, are expected to be linear in $\epsilon$.
In a perturbation expansion in powers of $\epsilon$,
the components $U_+(1)$ and $U_-(N)$ are {\em vanishing}
to dominant order in $\epsilon$, corresponding to
absorbing boundary conditions.
The first and last equations in (\ref{syst_pbcspecial})
can be considered as {\em new boundary conditions}
replacing (\ref{scatbc}), and the
remaining $2(N-1)$ equations can be written in matrix form as
\begin{equation}
\Lambda U = W^o U + \epsilon \Delta \left( U/\epsilon \right)
\label{wholeeq}
\end{equation}
where $W^o(k,i|k^\prime,j)$ is the matrix
representation of (\ref{scat_eig_eq}) with
$R_k = R_{k-1} = 1$ for the ABC case.
Inspection of (\ref{syst_pbcspecial}) shows that the perturbation
matrix $\Delta$ has the form
\begin{eqnarray}
\Delta (k,i|k^\prime,j) & = & a \delta(k,2)\delta(i,+)
\delta(k^\prime,1)\delta(j,+)\nonumber \\
& & + a \delta(k,N-1)\delta(i,-)\delta(k^\prime,N)
\delta(j,-),
\label{Deltamatrix}
\end{eqnarray}
where $\delta(k,l)$ is a Kronecker delta.
The eigenvalue equation (\ref{wholeeq}) can be solved
by a perturbation expansion around the solutions
$\{\Lambda_o, U^o_k \}$ of the {\em open}
system, discussed in subsection \ref{sect_PRWconf}, i.e.
$\Lambda = \Lambda_o + \epsilon \Lambda_1 + \cdots$
and $U_i = U_i^o + \epsilon U_i^1 + \cdots $,
yielding the equations of $\Or(1)$ and $\Or(\epsilon)$ :
\begin{eqnarray}
(\Lambda_o - W^o) U^o & = & 0 \nonumber \\
(\Lambda_o - W^o) U^1 + \Lambda_1 U^o & = & \Delta \; U^1,
\label{perturbation}
\end{eqnarray}
whereas the new boundary conditions follow from the first
and last line of (\ref{syst_pbcspecial}) :
\begin{eqnarray}
\Lambda_0 U^1_+(1) & = & a U_+^o(N) \nonumber \\
\Lambda_0 U^1_-(N) & = & a U_-^o(1).
\label{newbc}
\end{eqnarray}
Here $\Lambda_o$ and $U_i^o(k)$ are explicitly
given in (\ref{lambdaR}) and (\ref{ev_sin})
with $A=B=1$.
To solve the $\Or(\epsilon)$--equations in
(\ref{perturbation}), we need the left eigenvector $V_i^o(k)$
defined through $V^o W^o = \Lambda_o V^o$.
Using the symmetries in the explicit form of the
\mbox{$2(N-1)$--dimensional} matrix $W^o$,
it is straightforward to relate
the components of the left and right eigenvectors,
$V^o_{\pm}(k)$ and $U^o_{\pm}(k)$ respectively,
with the result :
\begin{eqnarray}
V_+^o(k) & = & U_-^o(k-1) = \sin q_o(N-k+1) \nonumber \\
V_-^o(k) & = & U_+^o(k+1) = \sin q_ok,
\label{newsin}
\end{eqnarray}
where $q_o = \pi / (N+a/b)$ as given in (\ref{qapprox}).
The eigenvalue in first order follows from
(\ref{perturbation}) by taking the inner product
of (\ref{perturbation}) with $V^o$, yielding
\begin{equation}
\epsilon \Lambda_1 = \epsilon
\frac{\langle V^o|\Delta|U^1 \rangle}{\langle V^o|U^o \rangle}
\simeq \frac{2}{\pi} \left( \frac{a}{b} \right)^2
(a+b)^{2-R} q_o^3 \;\;\;\;\;\;\;\;
(N \longrightarrow \infty)
\label{lambda1}
\end{equation}
where inner products are defined by
\begin{equation}
\langle X|Y \rangle = \sum_k \sum_{i=\pm 1} X_i(k) Y_i(k).
\end{equation}
Here numerator and denominator have been calculated from
(\ref{ev_sin}), (\ref{Deltamatrix}), (\ref{newbc}),
and (\ref{newsin}) with the result
\begin{eqnarray}
\langle V^o | \Delta | U^1 \rangle & = &
a V^o_+(2) U^1_+(1) + a V^o_-(N-1) U^1_-(N)
\nonumber \\
& & 2a^2 [ \sin q_o(N-1)]^2 / \Lambda_o
\simeq 2(a^2/b^2)(a+b) q_o^2
\nonumber \\
\langle V^o | U^o \rangle & = &
\cot q_o \sin q_o N - N \cos q_o N \nonumber \\
& \simeq & N + a/b = \pi/q_o.
\end{eqnarray}
We conclude that the largest eigenvalue $\Lambda(\beta)$
and corresponding Ruelle pressure $\psi(\beta) = \ln
\Lambda(\beta)$ in a PBC---configuration with
$N$ scatterers and a void of length $R$ decays
exponentially with a correlation length $\xi = 1/\ln(a+b)$
to the eigenvalue of a solid cluster of $N$ scatterers
with ABC's.

\begin{figure}
\centerline{\psfig{file=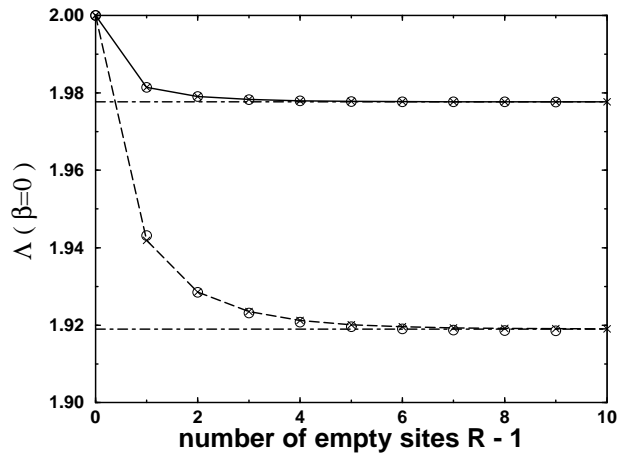,width=3.5in}}
\caption{\em Eigenvalue $\Lambda(\beta)$ at $\beta = 0$, for
a configuration with only one void containing $R-1$
empty sites and PBC, as a function of $R$.
The upper (lower) curves
correspond to the numerical results for
$N = 20$ ($N = 10$);
symbols (circles) give
the prediction based on equations
(\protect{\ref{Lambda38}})--(\protect{\ref{defq1}});
horizontal lines indicate the exact eigenvalue
$\Lambda_o(\beta=0)$ for
a PRW on a lattice of size $L^\prime=N$ with ABC.
For $\beta=0$, $\xi = 1/\ln 2 \simeq 1.4$ is
the correlation length.}
\label{specialPBC}
\end{figure}

In appendix A, we present a detailed calculation,
exact for all $R$ ($R = 1,2,\cdots$). It yields for the
largest eigenvalue
\begin{equation}
\Lambda(q) = (a+b) \{ 1-(a/2b) q^2 + \cdots \}
\label{Lambda38}
\end{equation}
where
\begin{equation}
q = q_1 = \pi/\left\{ N+ \frac{a}{b}
\left[ \frac{1+(a+b)^{1-R}}{1-(a+b)^{1-R}}
\right] \right\}.
\label{defq1}
\end{equation}
A plot of $\Lambda(q_1)$ is shown in Fig. \ref{specialPBC}.
The $\epsilon$--expansion of this result, with
$\epsilon = (a+b)^{1-R}$, agrees with the perturbation
result (\ref{lambda1}). For $R=1$ (no empty sites),
the resulting wave number reduces to $q_1=0$, as it
should for a closed system.

Again, equation (\ref{defq1}) shows that $q$ and $\Lambda(q)$
decay within a correlation length $\xi = 1/\ln(a+b)$
towards the corresponding values $q_o$ and $\Lambda_o(q_o)$
of an open system with a solid cluster of $N$ scatterers.

Equations (\ref{Lambda38}) and (\ref{defq1}) allow us
to calculate all chaos properties of the configuration
in figure (\ref{fig_conf},b) with the help of equation
(\ref{a11}).

\subsection{ABC---configurations with a void}

\label{sect_ABCvoid}

We consider an {\em open} system with two solid clusters
(see figure (\ref{fig_conf},c)) with respectively
$M$ and $\overline{M} = N-M$ scatterers, separated
by a void of length $R$ and study
the eigenvalue problem for $\beta < d_H$
so that $\Lambda(\beta) > 1$.
We expect that for sufficiently
large $R$ the $M$-- and $\overline{M}$--blocks
become independent, and the components
$U_-(M)$ and $\overline{U}_+(M+1)$ of the eigenvector,
corresponding to ``entering'' velocities, are small
and can be treated as new ABC's in a perturbation
calculation. We therefore write the eigenvalue equation
(\ref{scat_eig_eq}) as
a set of $2(M + \overline{M}-4)$ coupled equations
for the scattering amplitudes
$\{ U_i(k) | \overline{U}_i(k) \}$
with $(k,i) \in \{ 1^-,2^+,2^-,\cdots,(M-1)^-,M^+
|(M+1)^-,(M+2)^+,\cdots (N-1)^+,(N-1)^-,N^+ \}$
of the general form :
\begin{equation}
\Lambda\left(\begin{array}{c} U\\ \overline{U} \end{array}
\right) = \left( \begin{array}{cc} W^o & 0 \\
0 & \overline{W}^o \end{array} \right)
\left(\begin{array}{c} U\\ \overline{U} \end{array} \right)
+ \epsilon \Delta \left(\begin{array}{c}
U/\epsilon \\ \overline{U}/\epsilon  \end{array} \right)
\label{scatt2clust}
\end{equation}
with ``absorbing'' boundary conditions for the $M$--
and $\overline{M}$--blocks in the form
\begin{eqnarray}
U_+(1) & = & 0 \nonumber \\
\Lambda^R U_-(M) & = & \epsilon b
\left[\overline{U}_+(M+1)/\epsilon\right]
+ a \overline{U}_-(M+1) \nonumber \\
\Lambda^R \overline{U}_+(M+1) & = & a U_+(M) + \epsilon b
\left[U_-(M)/\epsilon\right]\nonumber \\
\overline{U}_-(N) & = & 0.
\label{BC44}
\end{eqnarray}
The block matrices $W^o(M)$ and $\overline{W}^o(\overline{M})$
refer respectively to the $M$-- and $\overline{M}$--clusters,
and have the same form as $W^o(N)$ for the $N$--cluster
in equation (\ref{Deltamatrix}).
The matrix $\Delta$ connects the block matrices
to the ``entering'' states $U_-(M)$ and
$\overline{U}_+(M+1)$ :
\begin{eqnarray}
\Delta(ki|k^\prime j) & = & a \delta(k,M-1) \delta(i,-)
\delta(k^\prime,M) \delta(j,-)\\
& & +
a \delta(k,M+2) \delta(i,+) \delta(k^\prime,M+1) \delta(j,+).\nonumber
\end{eqnarray}
The boundary conditions (\ref{BC44}) couple the two
blocks.
These boundary equations show that $U_-(M)$ and
$\overline{M}_+(M+1)$ for large $R$ can indeed be treated
as small quantities of order $\epsilon = \Lambda^{1-R}$,
as in subsection \ref{sect_PBCvoid}.

The analysis of this problem is similar to that in
subsection \ref{sect_PBCvoid}.
The eigenvalue equation (\ref{scatt2clust}) can be
solved by an expansion in powers of $\epsilon$.

The eigenvalue problem to {\em zeroth} order in $\epsilon$
reduces to two decoupled equations for the two isolated
$M$-- and $\overline{M}$--clusters with ABC, reading
\begin{equation}
\Lambda_o\left(\begin{array}{c} U^o\\ \overline{U}^o \end{array}
\right) = \left( \begin{array}{cc} W^o & 0 \\
0 & \overline{W}^o \end{array} \right)
\left(\begin{array}{c} U^o\\ \overline{U}^o \end{array} \right)
\label{scatt2clust0order}
\end{equation}
For sufficiently large $M$ and $\overline{M}$ the largest
eigenvalues of the block matrices are
$\Lambda_o(q_o)$ and
$\Lambda_o(\overline{q}_o)$ (see equation
(\ref{Lambda38})).
These are also eigenvalues for the whole system
(\ref{scatt2clust0order}) with right and left eigenvectors :
\begin{eqnarray}
\{ U^o(q_o)\; |\; 0 \} \;\; \mbox{ and } \;\;
\{ V^o(q_o)\; |\; 0 \} \;\;\;\; & \mbox{with} &
q_o=\pi/[M+a/b] \nonumber \\
\{ 0\; |\; \overline{U}^o(\overline{q}_o) \} \;\; \mbox{ and } \;\;
\{ 0\; |\; \overline{V}^o(\overline{q}_o) \} \;\;\;\; & \mbox{with} &
\overline{q}_o=\pi/[\overline{M}+a/b].
\label{eig46}
\end{eqnarray}
The right and left eigenvectors in (\ref{eig46})
are again given by (\ref{ev_sin}) and
(\ref{newsin}) with $N$ replaced by
$M$ and $\overline{M}$ respectively.
If $M>\overline{M}$, then $\Lambda(q_o) >
\Lambda(\overline{q}_o)$ and $\Lambda_o \equiv
\Lambda(q_o)$ is the largest eigenvalue
with the corresponding eigenvectors
$\{U^o(q_o)\;|\;\overline{U}^o(q_o)\equiv 0\}$ and
$\{V^o(q_o)\;|\;\overline{V}^o(q_o)\equiv 0\}$.

To {\em linear} order in $\epsilon$, the
boundary conditions (\ref{BC44}) require for the
components $\{U^1(q_o)\;|\;\overline{U}^1(q_o)\}$ :
\begin{eqnarray}
U_+^1(1) & = & 0 \nonumber\\
U_-^1(M) & = & a \overline{U}^o_-(M+1)
/ \Lambda_o = 0 \nonumber \\
\overline{U}_+^1(M+1) & = & a U_+^o(M)
/ \Lambda_o \label{BC47} \\
\overline{U}_-^1(N) & = & 0. \nonumber
\end{eqnarray}
The last equality on the second line follows as all
components of $\overline{U}^o(q_o)$ are vanishing.
First order perturbation theory for the largest eigenvalue
yields
\begin{equation}
\epsilon \Lambda_1 = \epsilon \langle V^o | \Delta | U^1
\rangle = 0,
\end{equation}
as a consequence of (\ref{BC47}).

Second order perturbation theory yields a non--vanishing
result, proportional to $\epsilon^2 \simeq (a+b)^{2-2R}$.
Therefore the largest eigenvalue for the configuration
of figure (\ref{fig_conf},c) as a function of the width
of the void $R$ has the form
\begin{equation}
\Lambda(R) \simeq \Lambda_o
+ \mbox{Const} \,\times\, e^{-(R-1)/\xi}
\label{decay}
\end{equation}
with a correlation length $\xi = [2\ln (a+b)]^{-1}$,
a factor $2$ smaller than in the previous subsection.

\begin{figure}
\centerline{\psfig{file=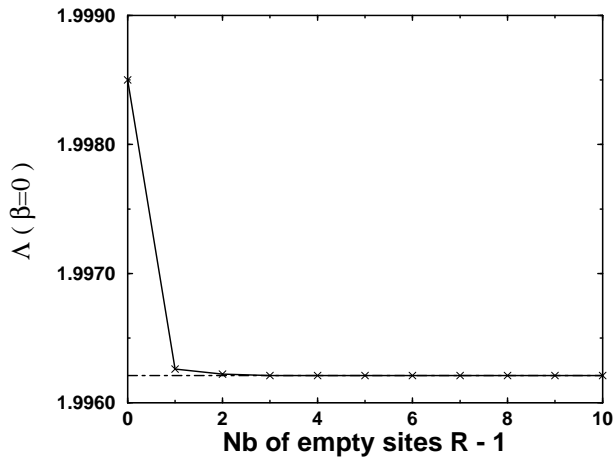,width=3.5in}}
\caption{\em Eigenvalue $\Lambda(\beta)$ at $\beta = 0$, for
a configuration with two clusters of sizes $50$ and $30$,
separated by a void of size $R$
and ABC.
The horizontal line indicates the exact eigenvalue
$\Lambda_o(\beta = 0) = 2$ for
a PRW on a lattice of size $L=50$ with ABC;
and $\xi(\beta=0) = 1/[2 \ln 2] \simeq 0.7$ is a theoretical
estimate for the correlation length, in good agreement
with the numerical results.}
\label{specialABC}
\end{figure}

In an ABC---configuration with two clusters containing
$M$ and $\overline{M}$ scatterers respectively, and
separated by a distance $R$, the largest eigenvalue
$\Lambda$ approaches the eigenvalue
$\Lambda_o \equiv \Lambda (q_o)$ given by
(\ref{Lambda38})--(\ref{defq1})
at an exponential rate.
This $\Lambda(q_o)$ is solely determined by the
{\em largest} cluster. The plot in Fig. \ref{specialABC}
shows the numerical solution of the eigenvalue problem.

The main conclusions of the previous subsections,
referring to $\beta < d_H$, can be directly generalized to
configurations with more clusters :

(i) The largest eigenvalue $\Lambda(\beta)$ in a
PBC---configuration (closed system) with at least
one void larger than $\xi$ is equal to the largest
eigenvalue for the corresponding ABC---configuration
(open system).

(ii) In any ABC--- or PBC---configuration with clusters
separated by distances $R_k > \xi$ ($k=0,1,\cdots,N$) the
largest eigenvalue is $\Lambda(q_o)$ given by (\ref{Lambda38})
with $q_o = \pi / [M_{max} + a/b]$, where $M_{max}$
is the number of scatterers in the largest solid cluster.

\begin{figure}
\centerline{\psfig{file=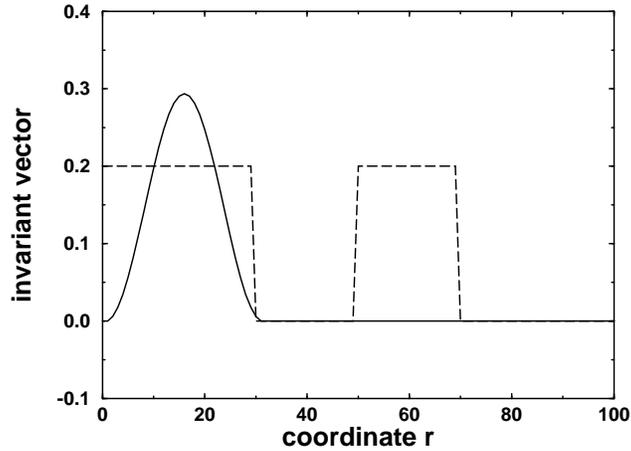,width=3.5in}}
\caption{\em Invariant vector $\pi_+(r)=u_+(r) v_+(r)$.
The location of the two clusters has been
indicated with dashed lines.}
\label{fig_ev1}
\end{figure}

In fact, we will use this last case (ii) to illustrate
the localization process mentioned in the introduction.
In section \ref{sect_models}, we have seen that the
largest eigenvalue $\Lambda(\beta)$ of the matrix $w(\beta)$
was dominating the dynamic partition function
$Z(\beta, t | x_o)$ at long times, and thus the Ruelle
pressure (equations (\ref{zlambdat})--(\ref{a10})).
On the other hand, we just found that, for $\beta<d_H$,
$\Lambda(\beta)$ is determined by the largest cluster
of the configuration, as if this cluster was isolated
and surrounded by absorbing boundaries.
It means that the trajectories which dominate in sum
(\ref{defZ}) are those which always remain localized
inside the largest cluster.
The other trajectories traveling throughout the system
may as well be omitted.
Thus the Ruelle pressure in large systems will not
reflect the global structure of the system but only characterize
the largest cluster present in the configuration.
It is precisely this phenomenon that will be referred to
as localization.
This can be illustrated by calculating the invariant
vector (see figure \ref{fig_ev1}). Only the states
corresponding to positive velocities are plotted here.
States with negative velocities give the same result,
up to a translation :
\begin{equation}
\pi_-(r) \langle u | v \rangle = 
u_-(r) v_-(r) = v_+(r+1) u_+(r+1)
= \pi_+(r) \langle u | v \rangle.
\end{equation}
After an infinite time, all the probability of finding the
particle is concentrated on the largest eigenvector.

To get some insight in how the Ruelle pressure varies with
the configuration of scatterers, we may compare the largest
eigenvalues $\Lambda(\beta)$ obtained for each of the three
configurations of figure \ref{fig_conf}, keeping a {\em fixed}
density $\rho = N/L = (M+\overline{M})/L$.
The eigenvalues are respectively
\begin{eqnarray}
\Lambda_a & = & (a+b)^\rho \left(
1-\frac{a}{2b} q_N^2 \right)^\rho \nonumber \\
\Lambda_b & = & (a+b) \left(
1-\frac{a}{2b} q_N^2 \right) \nonumber \\
\Lambda_c & = & (a+b) \left(
1-\frac{a}{2b} q_M^2 \right) \nonumber \\
\end{eqnarray}
As $\rho<1$ and $\beta<d_H$, i.e. $a+b>1$, it is
straightforward to show that
\begin{equation}
\Lambda_a < \Lambda_c < \Lambda_b.
\end{equation}
The largest eigenvalue is obtained when all scatterers
are packed together in one solid cluster, while the smallest
corresponds to the mean field configuration, where
no cluster is formed and thus localization is not possible.
In section \ref{sect_numerics}, it will appear that the largest
eigenvalue of any other configuration falls between
$\Lambda_a$ and $\Lambda_b$.
We will show that, for large $L$, most configurations
contain a largest cluster which will entirely
determine the Ruelle pressure.
Indeed, as we start to see with configuration (c),
localization is not specific to very special configurations,
but occurs more generally for most configurations.

Finally, we stress that, in this section, only the
case $\beta<d_H$ was considered.
For $\beta>>1$, a complementary phenomenon occurs,
i.e. localization in the largest void instead of largest
cluster \cite{appert96c}.

\section{Hausdorff dimension}

\label{sect_dh}

The eigenvalue equation in the representation 
(\ref{scat_eig_eq}) and the result of section
\ref{sect_PRWconf} enables us to carry out an exact
calculation of the Hausdorff dimension $d_H$ for
an open LLG, which is defined as the root of
$\psi(\beta = d_H) = 0$, or equivalently as the root
of $\Lambda(\beta=d_H) = 1$ on account of (\ref{a11}).
For a closed LLG there is no fractal repeller and $d_H=1$.

The important observation is that $d_H$ is {\em independent
of the quenched disorder}, and depends only on the total
number of scatterers. This can be seen by combining
equation (\ref{scat_eig_eq}) with the requirement
$\Lambda(\beta) = 1$.
The random variables $\{R_k\}$ disappear from the equation,
so that $d_H$ is the same as for the PRW in subsection
\ref{sect_PRWconf}.
It can be calculated by setting the right hand side of
(\ref{lambdaR}) equal to unity, and solving for $\beta$.
For large $N$ (i.e. small $q$), the root is
\begin{eqnarray}
\beta = d_H & = & 1-\left(\frac{p}{2q\lambda_o}\right)
\left(\frac{\pi}{N+a/b}\right)^2 + \Or (N^{-4})
\nonumber \\
& = & 1-\frac{D}{\lambda_{closed}}
\left(\frac{\pi}{L+a/(b\rho)}\right)^2 + \Or (L^{-4})
\label{dH}
\end{eqnarray}
where $D = p/2\rho q$ is the {\em exact}
diffusion coefficient of the one--dimensional
LLG \cite{vanbeijeren_s83},
and $\lambda_{closed} = \rho \lambda_o =
-\rho (p \ln p + q \ln q)$ is the exact Lyapunov exponent
for a closed LLG, as obtained in \cite{ernst95,dorfman_e_j95}.

\section{Numerical method}
\label{sect_numerics}

The remaining part of this paper describes the numerical
diagnostics, in which numerical and analytical results
will be compared. In this section we start with a description
of the numerical method, used to calculate the largest
eigenvalue of the large random matrix $w(\beta ;x|y)$ in
equation (\ref{defw}) for a fixed configuration of scatterers
characterized by a certain system size $L$ and
number of scatterers $N$.
Then the Ruelle pressure is obtained as the logarithm of
this eigenvalue (equation~(\ref{a11})).
 
In one dimension, a recurrence formula can be found
which allows us to compute numerically the exact value
of the determinant of $w(\beta) - \Lambda {\bf 1}$,
where ${\bf 1}$ is
the identity matrix. Then $\Lambda$ can
be determined as the largest root of the equation
 {\it det}$|w(\beta) - \Lambda {\bf 1}| = 0$r,
using Newton's method.
The recurrence formula is derived in appendix B.
This method can be applied provided that $L$ is not
too large (less than $400$).
Indeed, if the system size is larger,
numerical overflow problems occur.

For large system sizes ($L > 400$), and for $\beta \neq 1$,
the calculation of the determinant involves very large numbers that cannot be
handled by workstations. Under such circumstances, $\Lambda$
has been determined by using Arnoldi's algorithm, which
is an iterative method akin to Lanczos algorithm \cite{saad}.
Let $w$ be an $n\times n$ matrix whose largest eigenvalue
has to be determined.
The idea is to scan rapidly the eigenvector space, and
find a subspace ${\cal U}$ containing the $m$
most significant eigenvectors.
Then we compute an $m\times m$ matrix $H$ as a kind of
projection of $w$ onto the subspace ${\cal U}$.
The largest eigenvalue $\Lambda_H$ of $H$ associated
with the eigenvector $u_H$ yields an approximation
for the largest eigenvalue of $w$ and ${\cal U} u_H$ for the
corresponding eigenvector. If the result is not satisfactory,
the whole process is repeated, taking ${\cal U} u_H$
as an initial guess.
The method is explained with more details in Appendix C.
The size $m$ of the basis has to be tuned in order
to optimize the efficiency of the method.

\section{Random configurations.}
\label{sect_indiv}
 
In this section, we will illustrate that localization occurs not only
in the special configurations considered in section
\ref{sect_special}, but in the majority of random
configurations realized in large systems.

\begin{figure}
\centerline{\psfig{file=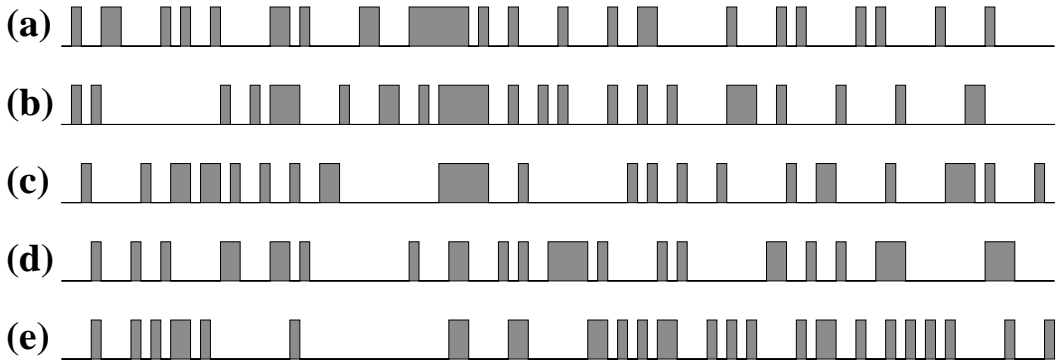,width=5.5in}}
\caption{\em
Random configurations obtained for $L=100$
and $\rho = 0.3$. The largest cluster is of size a) 6,
b) 5, c) 5, d) 4, e) 2.
The shaded areas represent the location of the scatterers
on the lattice.}
\label{specialconf}
\end{figure}

To show this we generate some random configurations
of a lattice of $L=100$ sites with a filling fraction
$\rho=0.3$, as shown in Figure \ref{specialconf}, and
calculate the largest eigenvalue $\Lambda(\beta)$
and corresponding Ruelle pressure
$\psi(\beta) = \ln \Lambda(\beta)$ numerically.
We further determine the size $M$ of the largest cluster
in each configuration, and calculate the largest
eigenvalue $\Lambda(\beta ; q_M)$ for an isolated
$M$--cluster, using equations (\ref{Lambda38})--(\ref{defq1})
with $q_M = \pi/[M+a/b]$.
The results are displayed in Table \ref{tab_specialconf}.
For comparison, the
mean--field value for the system size $L$ and
number of scatterers $N=\rho L$ is calculated
from equation (\ref{lambdaR}) and yields for $\beta=0$,
$\psi_{MF}(0) = 0.2078$ for ABC, and $\psi_{MF}(0) = 0.2079$
for PBC (where $q=0$ in equation (\ref{lambdaABC}).

The Ruelle pressure $\ln \Lambda(\beta)$ is in
fairly good agreement with the estimate
$\ln \Lambda(\beta ; q_M)$ for the three first
configurations (a)--(c) (see Table \ref{tab_specialconf}),
i.e. as soon as the largest cluster size $M$ is on the
order of $5$ sites.
In other words, the dynamic partition function (\ref{defZ})
and Ruelle pressure (\ref{a10}) calculated from the
subset of trajectories $\Omega_t$ that remain on the
largest cluster for all time, give already a fair
approximation to the actual Ruelle pressure defined
by summing in (\ref{defZ}) and (\ref{a10}) over all
trajectories $\Omega_t$ that stay inside the domain
${\mathcal D}$ for all time.
This means that there is already a large degree of
{\em localization} in configurations (a)--(c) and,
to a lesser extent, in (d) ---but not in (e)---, in spite
of the small system size $L=100$ and number $N=\rho L=30$
of scatterers.

On the other hand, if there is no large cluster
(as in configuration (e), where $M\le2$),
then $\Lambda(\beta ; q_M)$ is not a good estimate at all.
The mean field value is slightly better but still is a poor
estimate, as there are large fluctuations
in the distances between the scatterers.

\begin{table}
\begin{tabular}{cccc}
\\
\hline
configuration & Ruelle pressure & M & estimate \\
\hline
(a) & 0.59532 & 6 & 0.589 \\
(b) & 0.56275 & 5 & 0.549 \\
(c) & 0.55034 & 5 & 0.549 \\
(d) & 0.50916 & 4 & 0.481 \\
(e) & 0.35389 & 2 & 0.000 \\
\hline
\end{tabular}
\caption{\em For the random configurations of figure
\protect{\ref{specialconf}}, with $L=100$ and
$\rho=0.3$, we compare the
actual Ruelle pressure $\psi(\beta)$ with the estimate
$\ln \Lambda(\beta ; q_M)$
based on the largest cluster size $M$, for $\beta=0$.
PBC are used for configurations (a), (b), (c),
and ABC for (d), (e).
The corresponding mean--field values,
$\psi_{MF}(0) = 0.2079$ (PBC) and $\psi_{MF}(0) = 0.2078$ (ABC),
do not provide a sensible estimate.}
\label{tab_specialconf}
\end{table}

\subsection{Bounds on Ruelle pressure.}
\label{sect_bounds}

In \cite{appert96b,appert96c}, it has been shown that
the Ruelle pressure is bounded by
\begin{eqnarray}
\ln \Lambda(q_M) \simeq \ln (a+b) - \frac{a}{2b}\pi^2 / M^2
 \;\leq\; \psi_L (\beta ,
\rho ) &\leq \ln (a+b) \qquad & (\beta < 1) \nonumber \\
\beta (\ln q) \; 1/ \overline{M}
\;\leq \; \psi_L (\beta , \rho ) &\leq \;\; 0
\qquad & ( \beta > 1), \nonumber \\
\label{bounds}
\end{eqnarray}
where $M$ is the size of the largest cluster
and $\overline{M}$ of the largest void.
The upper bound is a direct consequence from
the inequalities
\begin{eqnarray}
a+b > 1 \;\;\;\; & \mbox{if} & \beta<1\\
a+b < 1 \;\;\;\; & \mbox{if} & \beta>1,
\end{eqnarray}
and we refer to \cite{appert96b,appert96c} for more details.
For $\beta<1$ (respectively $\beta>1$)
the lower bound is the Ruelle pressure obtained
by keeping only the largest cluster of the configuration
(or respectively the largest void, bordered by
2 scatterers) and using ABC.
In subsections \ref{sect_PBCvoid}, \ref{sect_ABCvoid},
and \ref{sect_indiv}, we have found
that for $\beta<1$, this is not only a lower bound
but also a good estimate for the Ruelle pressure,
as soon as $L$ is large enough.
A consequence is that, among all possible configurations,
the configuration with {\em all} scatterers solidly
packed in a single cluster gives the maximum value of the
Ruelle pressure.
On the other hand, the minimum
value is obtained for the mean field configuration
with scatterers spaced at regular intervals of length
$R=1/\rho$.

\begin{figure}
\centerline{\psfig{file=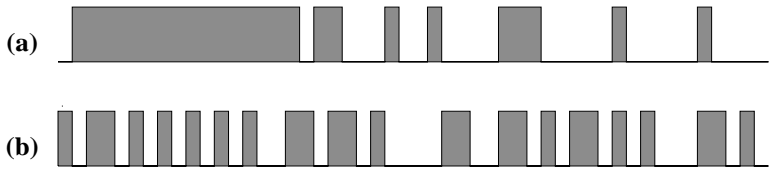,width=4in}}
\caption{\em Among $1000$ random configurations generated for
$L=50$, $N=25$, we select the configuration corresponding
to the largest (a) and smallest (b) $\Lambda$--value,
as determined numerically. The shaded areas correspond to the
location of scatterers.}
\label{profconf}
\end{figure}

\begin{figure}
\centerline{\psfig{file=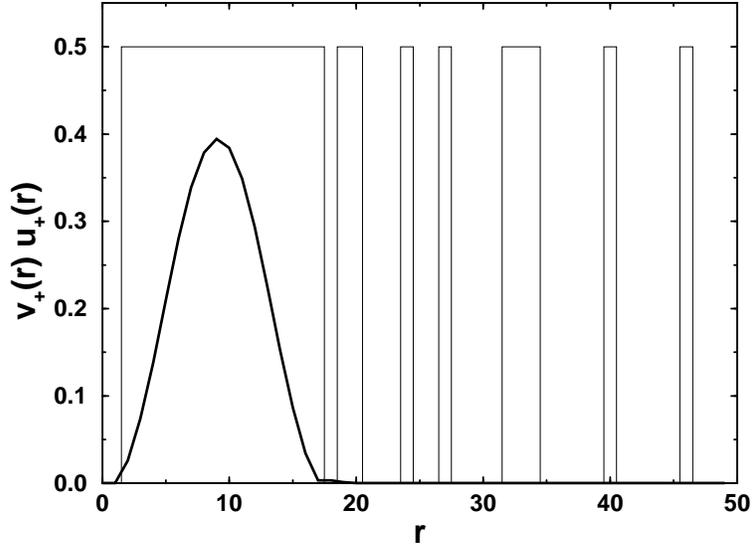,width=4in}}
\caption{\em Invariant measure $\pi(x)$ of Eq. (\ref{defpi})
for configuration (a) of
figure \protect{\ref{profconf}}. A thin line indicates the
profile for the location of scatterers.}
\label{figgevuv}
\end{figure}

\begin{figure}
\centerline{\psfig{file=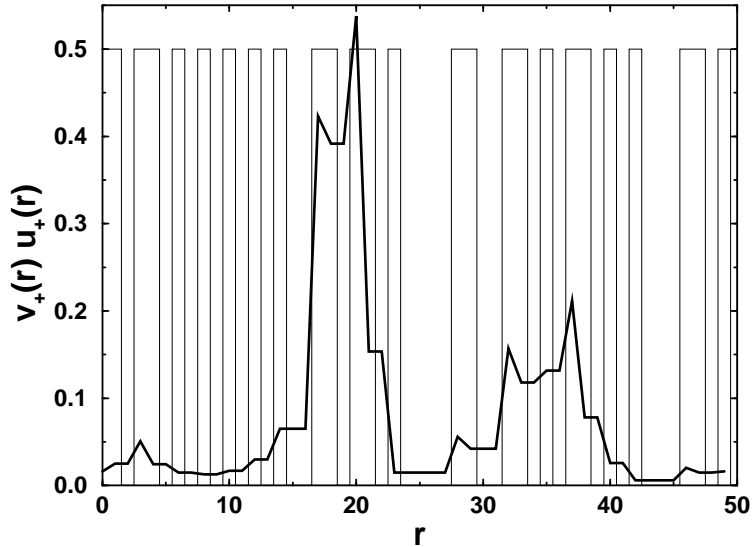,width=4in}}
\caption{\em Invariant vector for configuration (b) of
figure \protect{\ref{profconf}}. A thin line indicates the
profile for the location of scatterers.}
\label{figsevuv}
\end{figure}

As a confirmation, we have verified
(see fig. \ref{profconf}) that in a
given set of $1000$ configurations, the largest eigenvalue
for $\beta<1$
did correspond to a configuration where all scatterers
are essentially packed together, whereas the lowest value
was obtained for an ``almost mean-field'' configuration,
i.e. the distance between scatterers is more or less
constant. This is in agreement with our expectations
\cite{appert96c}.
To illustrate localization in these configurations,
we have plotted the invariant vector
$v_+(r) u_+(r)$ as a function of $r$
(figures \ref{figgevuv} and \ref{figsevuv}).
For configuration (a), the eigenvector is entirely
{\em localized} on the large cluster on the left.
Configuration (b), which is more of mean-field type,
correspond more or less to an {\em extended} state.
However, there is still partial localization in regions
of higher than average density.
 
\section{Distribution of largest cluster size}
\label{sect_largestsize}
 
Up till now we have studied single configurations.
In the next sections we will
present results averaged over the disorder and compare
them with the upper and lower bounds in Eq. (\ref{bounds}).
To do so, we need to average equation (\ref{bounds})
over all possible configurations of scatterers
with a fixed density $\rho$ (or a fixed $N$) and
a fixed system size $L$.
In order to determine $\langle 1/M^2 \rangle_\rho$
and $\langle 1/\overline{M} \rangle_\rho$,
we have used three different estimates for the distribution
of the largest cluster size.
Note that the distribution for the size of the largest
void is the same, upon exchanging scattering sites and
empty sites.
The first estimate, which is the most crude one, is just
the distribution for having
{\em at least} one cluster of size $M$ given by
\begin{equation}
P_1(M) = L \frac{\left(\begin{array}{c}
L-M-2 \\ N-M \end{array} \right)}{
\left( \begin{array}{c}
L \\ N \end{array}
\right)
}
\label{defP_1}
\end{equation}
This expression is valid only for large $M$ and for periodic
boundary conditions.
The numerator in (\ref{defP_1}) represents the number
of ways one can distribute
the $N-M$ remaining scatterers among the $L-M-2$
remaining empty sites, once a cluster of size
$M$ limited by two empty border sites has been put
in one of the $L$ possible locations.
The denominator is the total number of possible configurations.
 
The calculation of the distribution of largest cluster sizes
can be improved in the
following way: Let $A(M)$ be the fraction of realizations with
no cluster larger than $M$. Then
\begin{equation}
A(M-1) = A(M)\times [1 -  P_1(M)].
\end{equation}
Note that $1-P_1(M)$ is the probability that there is
{\em no} cluster of size $M$.
This recursion relation can be solved by iteration
starting at $M=N$, where $A(N)=1$. The result is :
\begin{equation}
A(M) = \Pi^N_{m=M+1} \left( 1-P_1(m) \right).
\end{equation}
The probability that the largest cluster size is $M$
is then
\begin{equation}
P_2(M) = A(M) - A(M-1) =
P_1(M) \Pi^N_{m=M+1} \left( 1-P_1(m) \right).
\label{defP_2}
\end{equation}
This expression for $P_2(M)$ can be calculated
numerically for each system size, by using the factorial expression
given above for $P_1(M)$.
This is the second expression which was used
for the numerical evaluation of lower bounds.
 
The third distribution $P_3(M)$ for the largest cluster
considered here was obtained by simple
generating a large number of configurations
and finding the largest cluster for each of them.
 
\begin{figure}
\centerline{\psfig{file=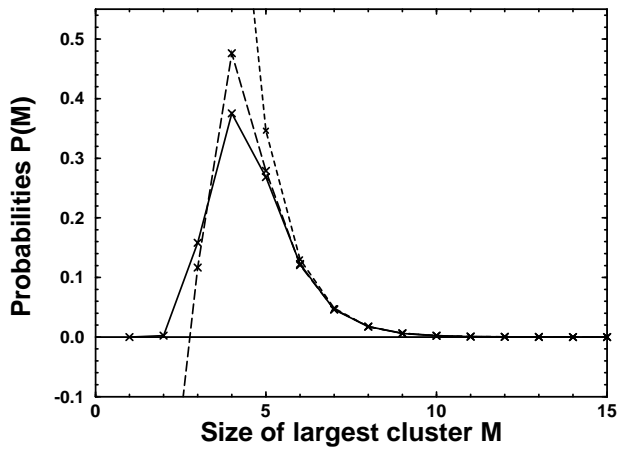,width=3.5in}}
\caption{\em Probability distribution for the largest
cluster size $M$ in a system of size $L=100$, and $\rho=0.4$.
The dotted, dashed, and solid lines correspond respectively
to $P_1(M)$, $P_2(M)$, and the direct measurement $P_3(M)$
obtained by generating 20 000 random configurations.}
\label{fig3distm}
\end{figure}

Figure \ref{fig3distm} compares these three
estimates for the distribution when $\rho = 0.4$, and
$L=100$.  As $P_1(M)$ is not bounded, it is cut off such that
the probability is normalized. The distribution
$P_2(M)$ has also to be cut off, otherwise it oscillates
between unphysical positive and negative values
at small $M$ (but we stress again that
the formulas (\ref{defP_1}) and (\ref{defP_2})
are not valid for small $M$--values).
The third distribution
$P_3(M)$ has been averaged over $20 000$ configurations.
 
\begin{figure}
\centerline{\psfig{file=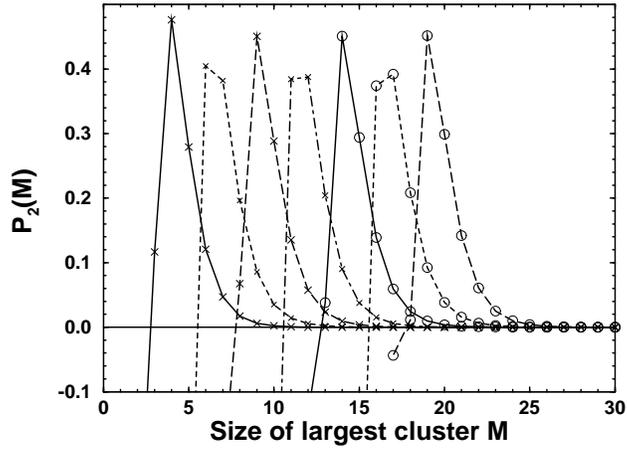,width=3.5in}}
\caption{\em Probability distribution $P_2(M)$
for the largest cluster size $M$ for $\rho=0.4$, $\beta=0$,
PBC, and for system sizes
$L$ increasing geometrically from $100$ to $10^8$.}
\label{figdistmL}
\end{figure}

\begin{figure}
\centerline{\psfig{file=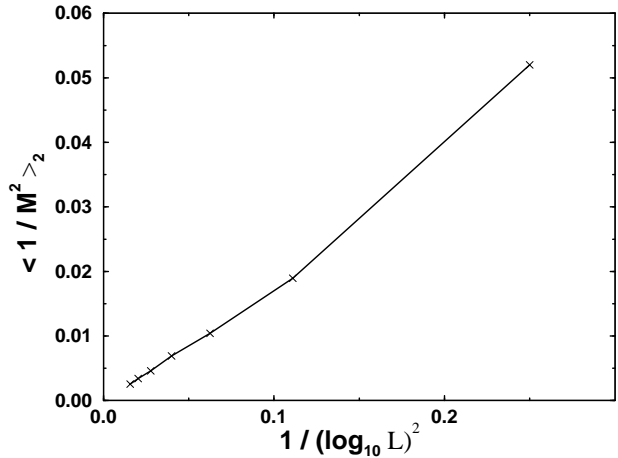,width=3.5in}}
\caption{\em Scaling properties of the second moment
$\langle 1/M^2 \rangle_2$, calculated from figure
\ref{figdistmL}, as a function of $1/(\log_{10}L)^2$ for
$L=100$ to $10^8$, at $\rho=0.4$ and $\beta=0$.}
\label{figscal1overM2}
\end{figure}

We now verify that the largest cluster size grows
as $\ln L$ for large $L$. Figure \ref{figdistmL} shows $P_2(M)$
for increasing system sizes. The system size is varied
from $100$ to $10^8$ and is multiplied by $10$ between
each successive estimation.
We check that each time
the size $L$ is multiplied by $10$, the maximum of the
distribution is shifted to the right by a constant value.
Another verification is made by plotting the
second moment $\langle 1/M^2 \rangle_2$ calculated
with the $P_2$--distribution
as a function of $(\log_{10}L)^{-2}$
(see figure \ref{figscal1overM2}).
 
\begin{figure}
\centerline{\psfig{file=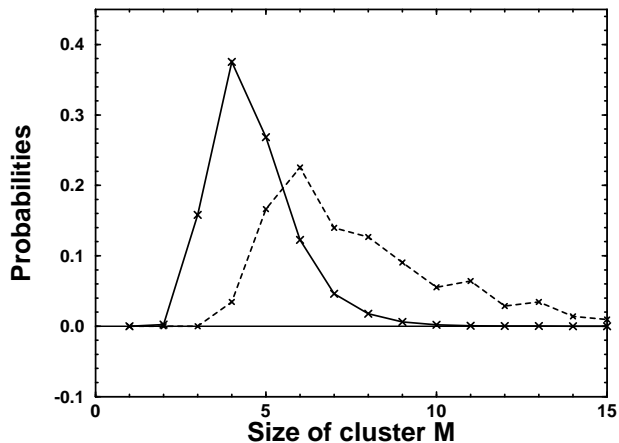,width=3.5in}}
\caption{\em Probability distribution for the largest
(solid line) and dominant (dashed line) cluster size
in a system of size $L = 100$, for $\rho = 0.4$
and $\beta = 0$.
These distributions were obtained by generating 20 000
random configurations.}
\label{fig_dominant}
\end{figure}

As already discussed in \cite{appert96c}, it is possible
in the one--dimensional case for $\beta<1$ and $L$ not too large,
that the Ruelle pressure is not determined
by the largest cluster but by a ``dominant''
cluster with average density $\rho + \Delta \rho$
intermediate between $\rho$ and $1$.
This is indeed what is observed in numerical simulations.
At $\beta = 0$ and for $L=100$, we have taken
all segments of all lengths, measured
the average density on each of them,
calculated the corresponding
Ruelle pressure using the mean field expression
(\ref{lambdaMFII}), and kept the one which gives the largest
value.
We call it the dominant cluster.
For some configurations it coincides with the largest
cluster, but not always.
Figure \ref{fig_dominant} compares the distributions
for dominant and largest clusters.
They are different, the first one
being slightly shifted towards larger values.
Note that a cluster, dominant for a given $\beta$, may not
be dominant for another $\beta$ value.

Another illustration of partial localization
in high density
regions instead of solid clusters was given by figure
\ref{figsevuv}.

\section{Average Ruelle pressure.}
\label{sect_average}
 
Now we can use any of the three estimates for the distribution
to calculate the
average (\ref{bounds}) over all possible configurations
and compare the resulting lower and upper bounds
with the numerical measurements of the Ruelle pressure.
For fixed $L$ and $\rho$, a large number of scatterer
configurations has been generated, and for each of them
the largest eigenvalue has been calculated. The average
of its logarithm yields a numerical value for the
Ruelle pressure.
 
\begin{figure}
\centerline{\psfig{file=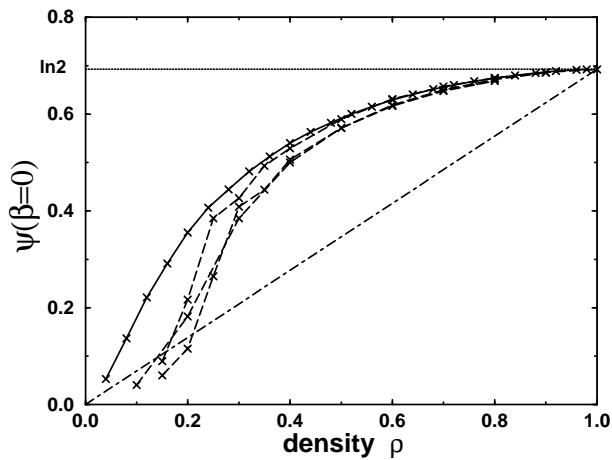,width=3.5in}}
\caption{\em Ruelle pressure $\psi(\beta = 0)$
(solid line) for $L=100$,
as a function of the density $\rho$,
compared with upper (dotted line) and lower bounds (dashed
lines). For comparison, the mean field prediction has also
been indicated (dashed--dotted line). }
\label{fignumthb0L100}
\end{figure}

We first consider the case where $\beta < 1$.
We have collected data for $\beta = 0$, for which
the Ruelle pressure equals the topological
entropy.
Figure \ref{fignumthb0L100} confirms that numerical
data are between the upper and lower bounds.
The lower bounds based on any of these three distributions
are qualitatively the same.
 
The estimate for the largest cluster distribution,
especially the one based on $P_1(M)$, may seem rather
crude. In fact, most of the system sizes that we
are able to explore numerically are too small
for the Ruelle pressure to be
entirely dominated by the largest cluster,
and no improvement of the estimates for $P(M)$
is likely to make the quantitative agreement better.
 
\begin{figure}
\centerline{\psfig{file=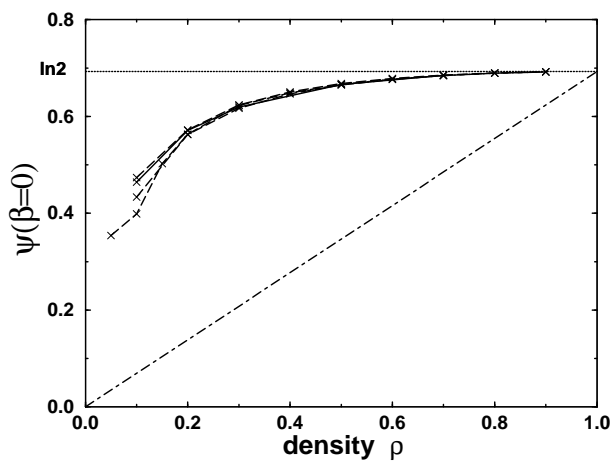,width=3.5in}}
\caption{\em Ruelle pressure $\psi(\beta = 0)$
(solid line) for $L=10000$,
as a function of the density $\rho$,
compared with upper (dotted line) and lower bounds (dashed
lines). For comparison, the mean field prediction has also
been indicated (dashed--dotted line). }
\label{fignumthb0L10000}
\end{figure}

However, when the system size is increased ($L = 10 000$),
we check --within the precision of our measurements-- that
the lower bound based on the largest cluster approximation
is indeed a good estimate of the Ruelle pressure
(figure \ref{fignumthb0L10000}).
It should be noted that at $\beta = 0$, the Ruelle
pressure is independent of $p$ and $q$,
and thus these results are valid both for large
or small $q$.
 
\begin{figure}
\centerline{\psfig{file=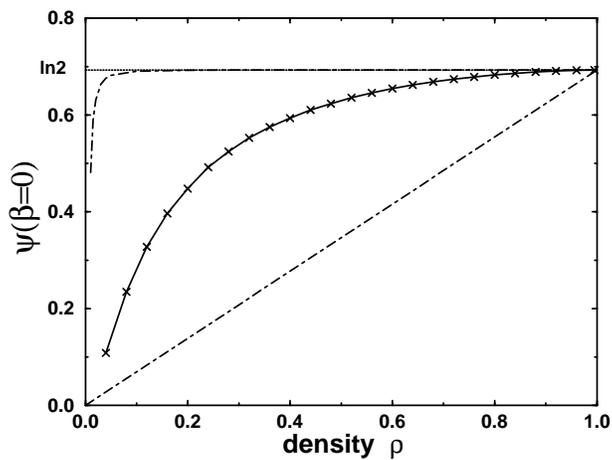,width=3.5in}}
\caption{\em Ruelle pressure $\psi(\beta=0)$
as a function of the density $\rho$ for $L=400$.
The solid line is an average over $10 000$
configurations. The dashed--dotted lines are the extreme
values obtained for specific configurations,
namely the `mean field' configuration
and the one where all scatterers are packed together in
a solid cluster. The upper bound
$\psi(0) = \ln 2$ is also indicated.}
\label{figextremeconf}
\end{figure}

One can also verify that for a given system size,
the mean field prediction gives a value much lower
than the average, whereas the configuration with
all scatterers are solidly packed together has a Ruelle
pressure almost equal to the upper bound.
This is illustrated in figure \ref{figextremeconf}
for $L=400$.
Thus for a configuration with $L=400$ and a given
density, any value between the two dashed--dotted
lines of figure \ref{figextremeconf} can be realized.
 
\begin{figure}
\centerline{\psfig{file=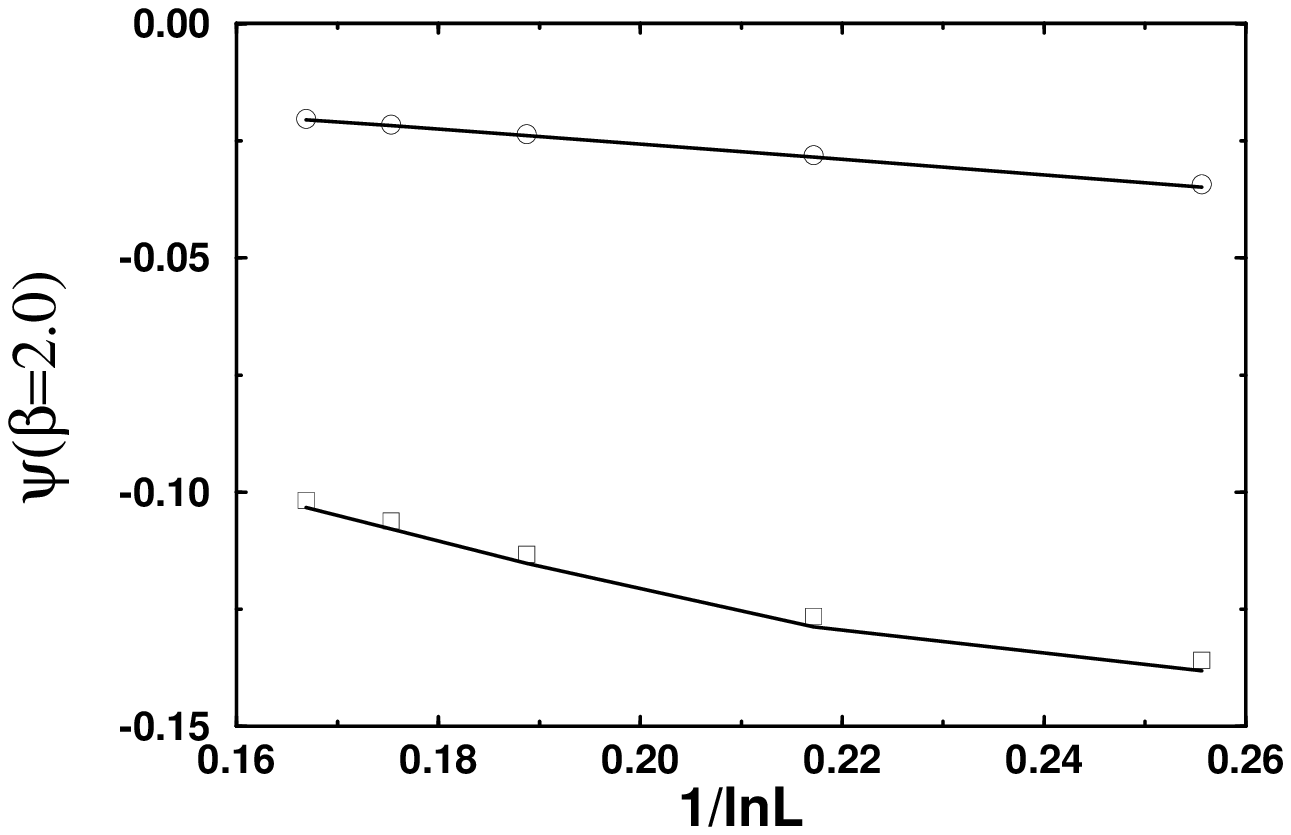,width=3.5in}}
\caption{\em Ruelle pressure as a function of the system size
in the case of strong backscattering ($p=0.2$) for
$\rho =0.2$ ($\circ$) or $\rho=0.8$ ($\Box$).
Lower bounds (solid lines) were obtained using
a numerical determination of the
largest cluster size distribution.}
\label{figpsmall}
\end{figure}

Now we address the case $\beta > 1$.
Tables \ref{tab1} and \ref{tab2} give some numerical values
for the measured or estimated Ruelle pressure,
and for the mean field prediction (\ref{lambdaMFII}),
in the case of $L=100$ and $\beta = 2$.
When $q$ is large (table \ref{tab1}), the lower bound
is not only a lower bound but also a good estimate of
the Ruelle pressure, and is much lower than the mean-field
value.
The numerical data are also displayed in figure
\ref{figpsmall}.
 
\begin{figure}
\centerline{\psfig{file=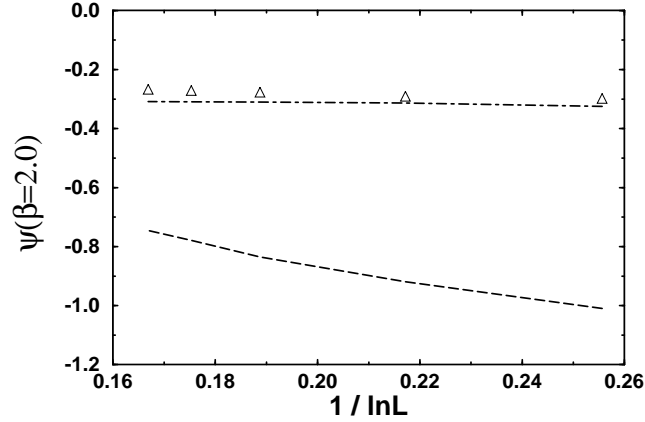,width=3.5in}}
\caption{\em Ruelle pressure as a function of the system size
in the case of weak backscattering ($p=0.8$)
at a high density $\rho=0.8$ ($\triangle$).
The lower bound (dashed line) was obtained using
a numerical determination of the
largest cluster size distribution.
The mean field prediction (dashed--dotted line) is a much
better estimate for the Ruelle pressure.}
\label{figplarge1}
\end{figure}

\begin{figure}
\centerline{\psfig{file=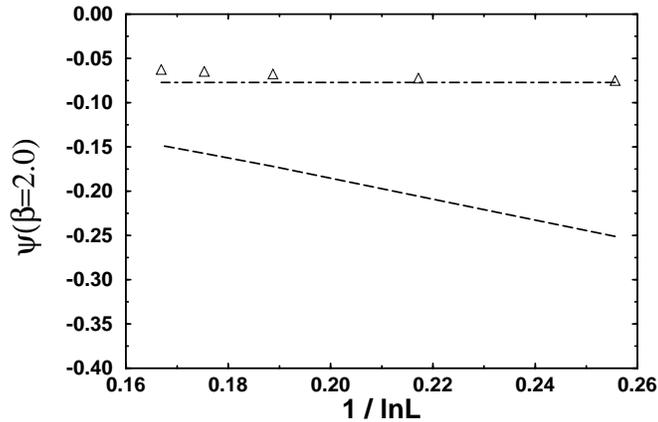,width=3.5in}}
\caption{\em Ruelle pressure as a function of the system size
in the case of weak backscattering probability ($p=0.8$)
at a low density $\rho=0.2$ ($\triangle$).
Again, the mean field prediction (dashed--dotted line) is a much
better estimate for the Ruelle pressure.}
\label{figplarge2}
\end{figure}

When $q$ is small (table \ref{tab2}, figure \ref{figplarge1}),
the situation is reversed. A good estimate is obtained
by using the mean field theory result, while the theoretical
lower bound differs significantly from the measured
Ruelle pressure.
We may roughly estimate under which condition
the lower bound will be a better estimate than
the mean field theory by comparing the corresponding
eigenvalues in (\ref{bounds}) and (\ref{lambdaR}) with
$R=1/\rho$ :
\begin{equation}
q^{\beta/\overline{M}} < (a+b)^\rho
\end{equation}
or with $b=q^\beta$
\begin{equation}
\overline{M} > \frac{|\ln b \,|}{\rho |\ln (a+b)\, |}.
\end{equation}
If $p>q$, then for $\beta>>1$ we have $a>>b$.
Localization will dominate over mean field if
\begin{equation}
\overline{M} > \frac{|\ln b|}{\rho |\ln a|}
= \frac{|\ln q|}{\rho |\ln p|},
\end{equation}
i.e. the system size $L$ must be typically larger
than $\exp (1/\rho \; \ln q / \ln p)$.
On the other hand, at fixed $L$, $\beta$ must be larger
than
\begin{equation}
\beta > 1 + \frac{(\ln q) |\ln (1-\rho)|}{(p\ln p + q\ln q)
\rho M}
\end{equation}
where $M \sim \ln L$.
This results from an expansion in powers of $\beta-1$
\cite{appert96c}.
It appears that, if $q$ is small, it is necessary
to go to very large $L$ and/or $\beta$ values to see
the occurrence of localization.
In the case of figure \ref{figplarge1},
$\beta = 2.0$ and $L = 400$
are not large enough to see this phenomenon.
Figure \ref{figplarge2} shows that this is
also the case for a different density $\rho$.

\begin{table}
\begin{tabular}{llll}
\\
\hline
$\rho$          & 0.2     & 0.5     & 0.8 \\
Numerical value & 0.02809 & 0.06583 & 0.12211 \\
Lower bound     & 0.02851 & 0.06682 & 0.12406 \\
Mean field      & 0.07713 & 0.19283 & 0.30853 \\
\hline
\end{tabular}
\caption{\em For $p = 0.2$ (large backscattering
probability),
the Ruelle pressure, evaluated numerically by averaging
over 10 000 configurations, is compared to the lower
bound and the mean field value, for different densities
of scatterers and a system size $L = 100$.}
\label{tab1}
\end{table}
 
\begin{table}
\begin{tabular}{llll}
\\
\hline
$\rho$          & 0.2     & 0.5     & 0.8 \\
Numerical value & 0.07218 & 0.17361 & 0.29033 \\
Lower bound     & 0.20584 & 0.48190 & 0.91901 \\
Mean field      & 0.07713 & 0.19283 & 0.30853 \\
\hline
\end{tabular}
\caption{\em For $p = 0.8$ (weak backscattering probability),
the Ruelle pressure, evaluated numerically by averaging
over 10 000 configurations, is compared to the lower
bound and the mean field value, for different densities
of scatterers and a system size $L = 100$.}
\label{tab2}
\end{table}

\begin{figure}
\centerline{\psfig{file=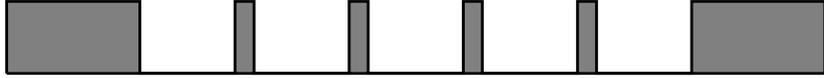,width=4.5in}}
\caption{Special configuration, in which
a trajectory exploring the whole low density region
may have a higher weight if $p>q$ than a trajectory remaining
in one of the voids.}
\label{figvoid}
\end{figure}

The picture of what happens when $p>q$
can still be refined.
The crossover between trajectories extended over the
whole lattice (mean field) and trajectories localized
in the largest void (lower bound) is in fact not
direct.
Some intermediate semi--localization may occur.
As mentioned in section \ref{sect_largestsize},
this was already evidenced for one--dimensional
systems at $\beta<1$ (and for all $p,q$ values)
\cite{appert96c}.
We will show now that a similar phenomenon occurs
for $\beta>1$ when $p>q$.
We then have not only $a>>b$, but also $a<<1$ ($\beta>1$).
Thus (i) on the one hand the particle has a tendency
to escape from the void because back-scattering
is weak; (ii) on the other hand, as $a<<1$, the free
propagation is till favored over forward propagation
through a scatterer.
If $a \simeq 1$, the second effect is negligeable and
the mean field prediction will be appropriate,
as it is the case in figures \ref{figplarge1}
and \ref{figplarge2}.
If $a<<1$, the competition between the two effects
may eventually promote some intermediate configurations.
As an illustration, we will consider the configuration
of figure \ref{figvoid} :
the whole lattice is solidly filled with scatterers
except in the low density region of size $R$. In this
region, $n$
isolated scatterers are placed at equal distance
from each other.
We will compare the weights of a trajectory $T_1$
undergoing only free propagation and back--scattering,
i.e. confined in a void in a strict sense between
two isolated scatterers;
and a trajectory $T_2$ going through the $n$ isolated
scatterers and thus exploring the entire low density region
of size $R$.

During $t$ time steps, $T_1$ will undergo
$t/[R/(n+1)]$ back--scatterings,
while $T_2$ will undergo $t/R$ back--scatterings
and $tn/R$ forward--scatterings.
Thus $T_2$ has a higher weight than $T_1$ if
\begin{equation}
q^{\beta\frac{t}{R}}
p^{\beta\frac{tn}{R}}
> q^{\beta\frac{t(n+1)}{R}}
\end{equation}
or equivalently
$p>q$.
This shows that as soon as $p>q$, localization
may occur not in the largest void but in a large
low density region.
However, when $L$ increases, we speculate that it is
more and more likely to find a large void that will
nevertheless dominate the result.

\section{Conclusion}
 
We conclude this paper with a number of remarks:
 
1) We have shown that the Ruelle pressure of Lorentz lattice
gases
in the limit of infinite systems is completely determined
by rare fluctuations in the configuration of scatterers.
Thus it carries no information on the global
structure of the disorder.
Numerical studies allowed us to show that this also holds
for finite but large systems, for all but a small range of
$\beta$ values.
Then we were able to predict quantitatively
the Ruelle pressure for all $\beta$ except
for a small region around $\beta =1$.
 
As already discussed in \cite{appert96b},
these results can be immediately generalized
to a whole class of diffusive systems with
static disorder. The special case of a continuous
Lorentz gas has been briefly considered in \cite{appert96c}.
 
Localization phenomena in fact appear very often
in physics, as soon as there is some competition
between energetically favorable configurations and
entropic effects.
Some localization effects were already shown
in the framework of the thermodynamic formalism
for deterministic maps \cite{beck_s93} or multifractals
\cite{feigenbaum_p_t89}.
However, it is the first time that such effects are
evidenced in hard spheres systems as resulting from the
quenched disorder. Localization occurs
on the most extreme fluctuation of the disorder.
For infinitely large systems, this fluctuation
may be arbitrarily large, which allows for a
very pronounced effect.
 
2) The Ruelle pressure is a characteristic of the dynamics
of the system not only for the isolated $\beta$ values where 
a direct interpretation can be given (see section
\ref{sect_models}), but also as a {\em whole}.
In the same way, in a power spectrum,
only some of the points
may receive an individual interpretation,
but the whole structure of the spectrum is
interesting.
An open question is to know if enough information
has been kept in the region around $\beta=1$ where
delocalization does {\em not} occur,
in order to be able for example to reconstruct
the structure of the disorder from it.
To this respect, it would be interesting
to rescale the region
around $\beta=1$ as the system size increases,
in order to prevent it from shrinking to zero
in the thermodynamic limit.
The scaling of this region with the system size
has been estimated in \cite{appert96c}.
 
When localization occurs, the information contained
in the Ruelle pressure
concerns the properties of individual scatterers.
More precisely, the knowledge of the $\beta$ regions
in which one given type of scatterers will dominate
yields a measure of what could be
called the isotropy of the different scatterers.
 
3) Numerical studies have been performed only in one dimension.
Analytical results
showed that the localization occurring
in the thermodynamic limit and the extension of the
delocalized region around $\beta=1$
can be generalized to higher dimensions.
Localization occurs also for finite
but large systems. The only difference with
the one--dimensional case is that now the lower bound
may not be a good estimate for the Ruelle pressure
at finite size. Indeed, the lower bound chosen here
was based on localization in hypercubic domains. In fact, it may
happen on domains with much more general shapes.
 
\vskip 1cm

{\bf Acknowledgments:}
The numerical values for the Ruelle pressure in
figure \ref{fignumthb0L100} and table \ref{tab_specialconf}
were obtained by C. Bokel.
We thank J.R. Dorfman and H. van Beijeren
for a fruitful collaboration, and
H.A. van der Vorst and G.L.G. Sleijpen
for useful discussions about iterative methods.
C. Appert acknowledges support of the foundation
``Fundamenteel Onderzoek der Materie'' (FOM), which is
financially supported by the Dutch National Science
Foundation (NWO), and of the ``Centre National de la
Recherche Scientifique'' (CNRS).

\vskip 1 true cm
{\sc Appendix A}
\vskip 1 true cm

In the case of $N$ scatterers packed in a single cluster
in a system
of size $L=N+R-1$ with PBC, a perturbative calculation
for the largest eigenvalue $\Lambda$ has been presented
in section \ref{sect_PBCvoid}.
Here we calculate the largest eigenvalue directly
from the set of equations (\ref{syst_pbcspecial}).

First notice that the bulk equations in (\ref{syst_pbcspecial})
impose the relation (\ref{Lambda38}) between $\Lambda$
and the wave number $q$, as can be found in a similar
way as equation (\ref{lambdaABC}).
Now the wavenumber $q$ has to be determined from the
boundary equations
\begin{eqnarray}
\Lambda^R U_+(1) & = & a U_+(N) + b U_-(N)
\nonumber \\
\Lambda U_-(1) & = & b U_+(2) + a U_-(2)
\nonumber \\
\Lambda U_+(N) & = & a U_+(N-1) + b U_-(N-1)
\nonumber \\
\Lambda^R U_-(N) & = & b U_+(1) + a U_-(1)
\label{eqbc}
\end{eqnarray}
We search for a solution of the form
\begin{eqnarray}
U_+(k) = A \exp [\imath q(k-1)] + c.c.
\nonumber \\
U_-(k) = B \exp [\imath q(N-k)] + c.c.
\end{eqnarray}
By inserting this form into (\ref{eqbc}), we obtain
four equations
which determine A and B (complex numbers).
Non trivial solutions exist only if the determinant of
the system vanishes. This yields a transcendental equation
for $q$ which can be solved using the ansatz
\begin{equation}
q = \frac{\pi}{N-1+\delta}.
\end{equation}
This ansatz is based on the assumption that the present
case is similar to ABC, as soon as $R$ is large enough.
This will be correct only for $\beta<1$.

The determinant is expanded in powers of $q$.
Replacing $\Lambda$ by its expression (\ref{Lambda38})
in terms of $q$, we find:
\begin{eqnarray}
\mbox{Det} & = & -4\, q^2\, (\delta b - a + b)
(\Lambda^R+a+b)
\nonumber \\
&& \times \left[\delta \,b\,(\Lambda^R-a-b)
\,- \,(a+b)(\Lambda^R+a-b)\right].
\end{eqnarray}
Setting this determinant equal to zero yields
two solutions. We select the one that gives
the smallest $q$, i.e. the largest eigenvalue,
and find expression (\ref{defq1}).

\vskip 1 true cm
{\sc Appendix B :} Recurrence formula for the
determinant of $w$
\vskip 1 true cm

For a one--dimensional system, the transition matrix
has a form such that an exact recursion relation can
be found to calculate its determinant.
We will first illustrate it for ABC.
The transition matrix is of the form
\begin{equation}
w = \left( \begin{array}{ccccccccccc}
 \Lambda & b_2 & a_2 &&&&&&&&\\
 b_1 & \Lambda &&&&&&&&&\\
&& \Lambda & b_3 & a_3 &&&&&&\\
& a_2 & b_2 & \Lambda &&&&&&&\\
&&&& \Lambda & b_4 &&&&&\\
&&& a_3 & b_3 & \Lambda &&&&&\\
&&&&&&\cdots&&&& \\
&&&&&&& \Lambda & b_{L-1} & a_{L-1} & \\
&&&&&&& b_{L-2} & \Lambda && \\
&&&&&&&&& \Lambda & b_L  \\
&&&&&&&& a_{L-1} & b_{L-1} & \Lambda 
\end{array} \right)
\end{equation}
If the $k$-th scatterer is located in site $j$,
we define $D_k$ as the determinant of the $2j-1$
first lines and columns.
An auxiliary quantity $E_k$ is obtained from $D_k$
by removing the last column and the last but one line.
Then the following recursion relations hold :
\begin{eqnarray}
D_{k+1} & = & \Lambda^{2R_k} D_k - b E_{k+1} \nonumber\\
E_{k+1} & = & b D_k + a^2 E_k.
\label{rec}
\end{eqnarray}
Using the initial conditions
\begin{eqnarray}
D_1 & = & \Lambda^{1+2R_o} \nonumber\\
E_1 & = & 0,
\label{ci}
\end{eqnarray}
we can iterate the equations (\ref{rec}) and find
the determinant for the whole matrix $w$ as
\begin{equation}
\mbox{Det}(w) = \Lambda^{1+2R_N} D_N.
\end{equation}

The recursive calculation is performed numerically.
For $\beta=0$ (i.e. $\Lambda\simeq 2$), it is possible
to calculate the determinant for matrices
up to size $4(L-1)^2$, beyond which the method is spoiled
by numerical overflow problems.

In the case of PBC, the recursion relations are slightly
more complicated. The matrix $w$ is now of size $4L^2$
and reads
\begin{equation}
w = \left( \begin{array}{ccccccccccccc}
\Lambda &&&&&&&&&&& a_L & b_L\\
& \Lambda & b_2 & a_2 &&&&&&&&&\\
a_1 & b_1 & \Lambda &&&&&&&&&&\\
&&& \Lambda & b_3 & a_3 &&&&&&&\\
&& a_2 & b_2 & \Lambda &&&&&&&&\\
&&&&& \Lambda & b_4 &&&&&&\\
&&&& a_3 & b_3 & \Lambda &&&&&&\\
&&&&&&&\cdots&&&&& \\
&&&&&&&& \Lambda & b_{L-1} & a_{L-1} && \\
&&&&&&&& b_{L-2} & \Lambda &&& \\
&&&&&&&&&& \Lambda & b_L & a_L \\
&&&&&&&&& a_{L-1} & b_{L-1} & \Lambda & \\
b_1 & a_1 &&&&&&&&&&& \Lambda
\end{array}
\right)
\end{equation}
We choose the origin of the lattice such that
there is {\em no} scatterer in sites $1$ and $L$.
For PBC, this is always possible if there are 2 holes in a row
somewhere in the configuration, which will be the case
for all configurations if $\rho<0.5$.
Then $b_1 = b_L = 0$.

It can be shown that the determinant of $w$ is given by
\begin{equation}
\mbox{Det}(w) = \Lambda^{2R_N+1} D_N - b D^*_N
-a^2 E^*_N +2 \Lambda(-\Lambda)^{L-1} a^N,
\end{equation}
where $D_N$ and $E_N$, $D^*_N$ and $E^*_N$, are obtained
from the recurrence formula (\ref{rec}) with respectively
the initial conditions (\ref{ci}) and
\begin{eqnarray}
D^*_{1} & = & \Lambda^{2(R_o-1)}b_2 - b E^*_1 \nonumber\\
E^*_{1} & = & {b_2}^2 - {a_2}^2.
\label{ci2}
\end{eqnarray}

\vskip 1 true cm
{\sc Appendix C :} Arnoldi's method
\vskip 1 true cm

Consider an $n \times n$ matrix $A$.

We start from a guess $u_o$ for the right eigenvector.
In a classical power method, one would apply the
matrix $A$ repeatedly to this vector until it is aligned
with the eigenvector associated with the largest
eigenvalue.
With Arnoldi's method, we built a basis of $m$ vectors
that will span the vector--space more rapidly.
If $j$ vectors $\{u_o, u_1, \cdots, u_{j-1}\}$
of the basis have been obtained already,
$u_j$ is defined as follows.
First matrix $A$ is applied :
\begin{equation}
u^\prime_j = A u_{j-1}.
\end{equation}
Then $u^\prime_j$ is orthogonalized with respect to
the $j$ first vectors :
\begin{equation}
u^{\prime\prime}_j = u^\prime_j - \sum_{k=0}^{j-1}
\left[ u^{\prime}_j . u_k \right] u_k.
\end{equation}
Finally, $u_j$ is equal to the normalized $u^{\prime\prime}_j$.
This process is iterated until $m$ vectors are obtained.

A reduced $m \times m$ matrix $H$ is defined by
\begin{equation}
h_{ij} = u^{i-1} u^{j},
\end{equation}
such that
\begin{equation}
A = {\cal U} H {\cal U}^T + B,
\end{equation}
where ${\cal U}=\{u_o, u_1, \cdots, u_m\}$,
and $B$ is expected to be small.
Notice that the definition of $H$ implies that
$h_{j+1,j}$ is the norm of $u^{\prime\prime}_j$,
and $h_{ij} = 0$ if $i>j+1$.
As a consequence it is straightforward to calculate
the determinant of $H$, and thus to find its largest
eigenvalue $\Lambda$ by Newton's method (we know that it is smaller
than $a+b$) and the associated eigenvector $y_H$.
The first approximation for the largest eigenvalue of $w$
is taken to be $\Lambda$ associated with ${\cal U} h_H$.
If it is not satisfactory, the whole process is repeated,
taking ${\cal U} h_H$as a new initial guess.
The size $m$ of the basis has to be tuned in order to
optimize the efficiency of the method.

A difficulty that this method shares with other iterative
methods occurs when the largest eigenvalue is almost
degenerate with the next smaller one. Then we may by mistake
converge towards the second one. However, this is not
a real problem as long as we are interested in the
eigenvalue itself.

It should also be noted that, as usual,
convergence theorems exist only for symmetric matrices,
whereas the method has been applied here to non--symmetric
matrices.

\bibliographystyle{unsrt}

\begin{thebibliography}{10}

\bibitem{appert96c}
C.~Appert, H.~van Beijeren, M.H. Ernst, and J.R. Dorfman.
\newblock Thermodynamic formalism and localization in {L}orentz lattice gases.
\newblock {\em to appear in J. Stat. Phys}, vol. 87, 1996.

\bibitem{beck_s93}
C.~Beck and F.~Schl{\"{o}}gl.
\newblock {\em Thermodynamics of chaotic systems}.
\newblock Cambridge U. Press, 1993.

\bibitem{ruelle78}
David Ruelle.
\newblock {\em Thermodynamic Formalism}.
\newblock Addison Wesley Publ. Co. (Reading, Mass, 1978)., 1978.

\bibitem{feigenbaum_p_t89}
M.J. Feigenbaum, I.~Procaccia, and T.~T\'el.
\newblock Scaling properties of multifractals as an eigenvalue problem.
\newblock {\em Phys. Rev. A}, 39:5359, 1989.

\bibitem{gaspard_b95}
P.~Gaspard and F.~Baras.
\newblock Chaotic scattering and diffusion in the {L}orentz lattice gas.
\newblock {\em Phys. Rev. E}, 51:5332, 1995.

\bibitem{appert96b}
C.~Appert, H.~van Beijeren, M.H. Ernst, and J.R. Dorfman.
\newblock Thermodynamic formalism in the thermodynamic limit: Diffusive systems
  with static disorder.
\newblock {\em PRE}, 54:R1013, 1996.

\bibitem{gaspard_n90}
P.~Gaspard and G.~Nicolis.
\newblock {\em Phys. Rev. Lett.}, 65:1693, 1990.

\bibitem{ernst_d95}
M.H. Ernst and J.R. Dorfman.
\newblock Chaos in {L}orentz lattice gases.
\newblock In {J.J. Brey, J. Marro, J. M. Rubi, M. San Miguel editors}, editor,
  {\em 25 Years of Non-Equilibrium Statistical Mechanics}, pages 199--210.
  Springer Verlag, Berlin, 1995.

\bibitem{gaspard_d95}
P.~Gaspard and B.~Dorfman.
\newblock Chaotic scattering theory, thermodynamic formalism, and transport
  coefficients.
\newblock {\em Phys. Rev. E}, 52:3525, 1995.

\bibitem{ernst95}
M.H. Ernst, J.R.Dorfman, R.~Nix, and D.~Jacobs.
\newblock Mean field theory for {L}yapunov exponents and {K}olmogorov-{S}inai
  entropy in {L}orentz lattice gases.
\newblock {\em Phys. Rev. Lett.}, 74:4416, 1995.

\bibitem{vanbeijeren_s83}
H.~van Beijeren and H.~Spohn.
\newblock Transport properties of the one-dimensional stochastic lorentz model
  - i, velocity autocorrelation function.
\newblock {\em J. Stat. Phys.}, 31:231--254, 1983.

\bibitem{dorfman_e_j95}
J.R. Dorfman, M.H. Ernst, and D.~Jacobs.
\newblock Dynamical chaos in the {L}orentz lattice gas.
\newblock {\em J. Stat. Phys.}, 81:497--513, 1995.

\bibitem{saad}
Y.~Saad.
\newblock {\em Numerical methods for large eigenvalue problems}.
\newblock revised version (unpublished), 1995.

\end{thebibliography}

\end{document}